\documentclass[11pt,a4paper]{article}
\pdfoutput=1
\usepackage{jheppub}
\usepackage{subcaption}
\DeclareMathOperator{\Tr}{Tr}

\newcommand{\ri}{\mathrm{i}}

\newcommand{\cob}{\delta}

\newcommand{\hf}{\frac{1}{2}}
\newcommand{\qu}{\frac{1}{4}}
\newcommand{\til}[1]{\widetilde{#1}}

\renewcommand{\b}[1]{\overline{#1}}
\newcommand{\del}{\partial}

\newcommand{\bra}{\langle}
\newcommand{\ket}{\rangle}
\newcommand{\la}{\lambda}

\newcommand{\ka}{\kappa}
\newcommand{\h}[1]{\widehat{#1}}
\newcommand{\bt}{\beta}
\newcommand{\ga}{\gamma}

\newcommand{\al}{\alpha}

\newcommand{\rt}[1]{\sqrt{#1}}
\newcommand{\cO}{\mathcal{O}}

\newcommand{\cF}{\mathcal{F}}
\newcommand{\cR}{\mathcal{R}}
\newcommand{\cD}{\mathcal{D}}

\newcommand{\cL}{\mathcal{L}}
\newcommand{\cB}{\mathcal{B}}
\newcommand{\cA}{\mathcal{A}}

\newcommand{\cK}{\mathcal{K}}
\newcommand{\bbZ}{{\mathbb Z}}
\newcommand{\JT}{{\mbox{\scriptsize JT}}}
\newcommand{\eff}{{\mbox{\scriptsize eff}}}
\newcommand{\conn}{{\mbox{\scriptsize c}}}
\newcommand{\gs}{g_{\rm s}}
\newcommand{\tpartial}{\tilde{\partial}}
\newcommand{\hz}{\hat{z}}
\DeclareMathOperator{\Erf}{Erf}
\DeclareMathOperator{\Erfc}{Erfc}

\begin{document}

\title{Multi-boundary correlators in JT gravity}

\author[a]{Kazumi Okuyama}
\author[b]{and Kazuhiro Sakai}

\affiliation[a]{Department of Physics, Shinshu University,\\
3-1-1 Asahi, Matsumoto 390-8621, Japan}
\affiliation[b]{Institute of Physics, Meiji Gakuin University,\\
1518 Kamikurata-cho, Totsuka-ku, Yokohama 244-8539, Japan}

\emailAdd{kazumi@azusa.shinshu-u.ac.jp, kzhrsakai@gmail.com}

\abstract{
We continue the systematic study of the thermal partition function
of Jackiw-Teitelboim (JT) gravity started in [arXiv:1911.01659].
We generalize our analysis to the case of multi-boundary correlators
with the help of the boundary creation operator.
We clarify how the Korteweg-de Vries constraints arise
in the presence of multiple boundaries,
deriving differential equations obeyed by the correlators.
The differential equations allow us to
compute the genus expansion of the correlators
up to any order without ambiguity.
We also formulate a systematic method of calculating the WKB expansion
of the Baker-Akhiezer function and the 't Hooft expansion
of the multi-boundary correlators.
This new formalism is much more efficient than
our previous method based on the topological recursion.
We further investigate the low temperature expansion of
the two-boundary correlator.
We formulate a method of computing it up to any order
and also find a universal form of the two-boundary correlator
in terms of the error function.
Using this result we are able to write down the analytic form of
the spectral form factor in JT gravity
and show how the ramp and plateau behavior comes about.
We also study the Hartle-Hawking state
in the free boson/fermion representation of the tau-function
and discuss how it should be related to the multi-boundary correlators.
}

\maketitle

\section{Introduction}\label{sec:intro}

Jackiw-Teitelboim (JT) gravity \cite{Jackiw:1984je,Teitelboim:1983ux} is a very
useful
toy model to study various issues in quantum gravity and holography.
As discussed in \cite{Almheiri:2014cka,Maldacena:2016upp,Jensen:2016pah,Engelsoy:2016xyb},
JT gravity is holographically dual to the low energy Schwarzian sector of the
Sachdev-Ye-Kitaev (SYK) model \cite{Sachdev,kitaev2015simple}.
In a recent paper \cite{Saad:2019lba} 
Saad, Shenker and Stanford showed that the partition function
of JT gravity on asymptotically Euclidean AdS spacetime
is equal to the partition function of 
a certain double-scaled random matrix model
and the contributions of
higher genus spacetimes originated from the splitting and joining of baby universes
is captured by the $1/N$ expansion of the matrix model. See also
\cite{Stanford:2019vob,Blommaert:2019wfy,Okuyama:2019xbv,Johnson:2019eik,Kapec:2019ecr,Betzios:2020nry} for related works in this direction.
This opens up an interesting avenue to study
the effect of topology change in holography using the powerful techniques of the random matrix 
theory. This connection between JT gravity and the random matrix model
comes from the fact that the density of states in Schwarzian theory
is exactly equal to the planar (genus-zero) eigenvalue density of
the random matrix model which arises
in the topological recursion of the Weil-Petersson volume 
\cite{Eynard:2007kz}.
This connection is very interesting from the viewpoint of holography.
It clearly shows that JT gravity is dual to an ensemble
of boundary theories and the partition function
on asymptotic AdS spacetime with renormalized boundary length $\bt$ is
interpreted as the ensemble average $\bra Z(\bt)\ket=\bra \Tr e^{-\bt H}\ket$
over the random Hamiltonian $H$.

In our previous paper \cite{Okuyama:2019xbv}, we showed that
JT gravity is nothing but a special case of the
Witten-Kontsevich topological gravity \cite{Witten:1990hr,Kontsevich:1992ti}
and we studied the  partition function $\bra Z(\bt)\ket$ of JT gravity
on spacetime with a single asymptotic boundary in detail.
In particular, we found that $\bra Z(\bt)\ket$ is written as
the expectation value of the macroscopic loop operator in 2d gravity 
\cite{Banks:1989df}.
The important difference of JT gravity from the known example of 2d gravity is 
that infinitely many couplings $t_k$ are turned on
with a specific value $t_k=\ga_k$ with
\begin{equation}
\begin{aligned}
 \ga_0=\ga_1=0,\quad \ga_k=\frac{(-1)^k}{(k-1)!}~~~(k\geq2).
\end{aligned} 
\label{eq:gammavalue}
\end{equation}
By generalizing the approach of Zograf \cite{Zograf:2008wbe},
we found that the contributions of the higher genus
topologies can be systematically computed by
making use of the KdV constraint obeyed by the partition function. 
As emphasized in \cite{Zograf:2008wbe}, this method
serves as a very fast algorithm for the higher genus computation
compared to the  Mirzakhani's recursion relation for the Weil-Petersson volume
\cite{mirzakhani2007simple}.
We also found that in the low temperature regime the genus expansion
can be reorganized in the following scaling limit, which we call the 't Hooft limit
\begin{equation}
\begin{aligned}
 \hbar\to0,~\bt\to\infty\quad\text{with}\quad\la=\hbar\bt~~~\text{fixed},
\end{aligned} 
\label{eq:tHooft}
\end{equation}
where $\hbar\sim e^{-S_0}$ is the genus-counting parameter.
In this limit the free energy $\log \bra Z(\bt)\ket$
admits the 't Hooft expansion
\begin{equation}
\begin{aligned}
 \log \bra Z(\bt)\ket=\sum_{k=0}^\infty \hbar^{k-1}\cF_k(\la)
\end{aligned} 
\label{eq:CF-expand}
\end{equation}
and we found the analytic form of the first few terms of $\cF_k(\la)$.

We emphasize that this 't Hooft limit is not just taking the low temperature limit 
and replacing the Schwarzian density of states $\rho(E)\sim\sinh(2\rt{E})$ by the Airy
one $\rho(E)\sim \rt{E}$. Even after taking the 't Hooft limit, we still keep all the
non-trivial information of the spectral curve $y=\hf\sin(2\rt{x})$ 
of JT gravity matrix model.
In particular, the leading term $\cF_0(\la)$ in \eqref{eq:CF-expand}
is given by an integral on the spectral curve $y=\hf\sin(2\rt{x})$ 
\begin{equation}
\begin{aligned}
 \cF_0(\la)=2\int_0^{\la/2}xdy=\int_0^{\la/2} \hf \arcsin(2y)^2 dy.
\end{aligned} 
\end{equation}
The Airy case corresponds to the cubic polynomial 
$\cF_0(\la)_{\text{Airy}}=\frac{\la^3}{12}$ and we start to see the deviation from the Airy case at the order $\cO(\la^5)$. 
One might worry that by taking the 't Hooft limit we throw away all the interesting 
part of the black hole physics coming from the high energy states. 
However, as we will see in section
\ref{sec:SFF}, we indeed observe the Hawking-Page like transition between the disconnected 
Euclidean black holes and the connected Euclidean wormhole within the leading
approximation of the 't Hooft expansion. This clearly shows that
our 't Hooft expansion captures the interesting part of the physics of black holes.
Another concern is that there are only order one number of 
states left above the ground state 
after taking the 't Hooft limit and the naive gravity description breaks down. However,
JT gravity is dual to an averaged system with continuous density of states
from the beginning and it is not sensitive enough to distinguish black hole microstates. 
Moreover,
in the low temperature limit $\bt\to\infty$ the boundary has a macroscopic length 
in units of the Planck length and hence there is no problem in describing
such a situation by a smooth geometry. See also 
\cite{Iliesiu:2020qvm} for a recent discussion of the absence of mass gap in the spectrum
of the near-extremal charged black hole in 4d, whose near horizon dynamics is described by JT gravity.

In the present paper, we will study the partition function of JT gravity
on spacetimes with multiple boundaries
by generalizing the method of KdV equation in \cite{Okuyama:2019xbv}.
We find that the KdV constraints for the connected part of the multi-boundary
correlator $ \bra \prod_i Z(\bt_i)\ket_\conn$ is obtained
by acting the ``boundary creation operators'' $B(\bt_i)$
to the original KdV equation for the potential $u(x)$.
Our boundary creation operator $B(\bt)$ is the same as the
one discussed in the old 2d gravity literature \cite{Moore:1991ir,Ginsparg:1993is}
which is based on the idea that the macroscopic loop operator is expanded in terms of 
the microscopic loop operators in the limit $\bt\to0$, up to 
the so-called non-universal terms which scale with negative powers of $\bt$.
We can systematically compute the genus expansion of the correlator 
$\bra \prod_i Z(\bt_i)\ket_\conn$ by solving this KdV constraints recursively.
Most of the computation can be done away from the ``on-shell'' value of the couplings 
\eqref{eq:gammavalue}. In particular, we define the off-shell 
generalization of the effective potential and its Legendre transform, 
the off-shell free energy. We find that the multi-boundary correlators can be written 
in terms of a certain combination of the off-shell free energy in the 't Hooft limit.
We also study the WKB expansion of the Baker-Akhiezer (BA) 
functions.\footnote{
We will refer to 
the $\hbar$-expansion of a function of energy eigenvalue $\xi$
as ``the WKB expansion'' while the $\hbar$-expansion of a function of
the 't Hooft parameter $\la$ as ``the 't Hooft expansion.''
They are related by the saddle point approximation of the integral
such as \eqref{eq:nptint}.}

In this paper we will focus on the two-point function 
and compute its
genus expansion 
using the above formalism. We also compute its low temperature expansion and study 
its behavior in the 't Hooft limit as well.
It turns out that the two-point function in JT gravity
is expressed in terms of the error function,
which is a natural generalization of the known result
of pure topological gravity \cite{okounkov2002generating}.
From the bulk gravity viewpoint, the connected part of the
two-point function $\bra Z(\bt_1) Z(\bt_2)\ket_\conn$
corresponds to a Euclidean wormhole 
 (also known as the ``double trumpet''
\cite{Saad:2018bqo}) connecting
the two asymptotically AdS boundaries with renormalized lengths $\bt_1,\bt_2$.
The analytic continuation of the two-point function
$\bra Z(\bt+\ri t) Z(\bt-\ri t)\ket_\conn$, known as 
the spectral form factor (SFF), 
is of particular
interest in the context of quantum chaos
and the SFF is widely studied in the SYK model and JT gravity 
\cite{Garcia-Garcia:2016mno,Cotler:2016fpe,Saad:2018bqo,Saad:2019pqd}.
We find the analytic form of the SFF in the 't Hooft limit
and show that the SFF in JT gravity exhibits the characteristic feature of 
the so-called ramp and plateau, as expected for a chaotic system with random matrix statistics
of eigenvalues.

In a recent interesting paper \cite{Marolf:2020xie},
Marolf and Maxfield considered the boundary creation operators
in the context of the AdS/CFT correspondence
and made some interesting argument on the baby universe Hilbert space building upon
the earlier works by Coleman \cite{Coleman:1988cy} and
by Giddings and Strominger \cite{Giddings:1988cx,Giddings:1988wv}.
The argument in \cite{Marolf:2020xie} is mostly based on the intuition coming from
a simple toy model, which is not a full-fledged JT gravity.
It is interesting to ask how our boundary creation operator $B(\bt)$
fits into the story in \cite{Marolf:2020xie}, but we do not have a clear understanding
of it.
We make some preliminary remarks on this problem in section \ref{sec:boundary}
and leave the details for a future work.

This paper is organized as follows. In section \ref{sec:genus},
we compute the
genus expansion of multi-boundary correlators using the KdV constraint obeyed by these
correlators.
In section \ref{sec:wkb}, 
we consider the WKB expansion of the Baker-Akhiezer functions
and the 't Hooft expansion of the multi-boundary correlators.
Along the way, we define the off-shell extension of the effective potential and
the free energy.
In section \ref{sec:lowT}, we compute the low temperature expansion of the 
two-boundary correlator.
In section \ref{sec:SFF}, we study the spectral form factor
in JT gravity and show that it exhibits the ramp and plateau behavior
as expected for chaotic system.
In section \ref{sec:boundary}, we consider the
free boson/fermion representation of the
$\tau$-function and discuss the boundary creation operator in this formalism.
Finally we conclude in section \ref{sec:conclusion}.
In appendix \ref{sec:micro}, we consider the wavefunction of microscopic loop operators
in the 't Hooft limit.

\section{Genus expansion}\label{sec:genus}

\subsection{Basics and conventions}

In this paper we will generalize our method \cite{Okuyama:2019xbv}
developed for one-boundary partition function
to the case of multi-boundary correlators.
To begin with, let us summarize basics,
notations and conventions.

As we showed in \cite{Okuyama:2019xbv},
JT gravity can be regarded as a special case of
the general Witten-Kontsevich topological gravity
\cite{Witten:1990hr,Kontsevich:1992ti}.
In this model the intersection numbers
\begin{align}
\label{eq:intersec}
\langle\kappa^\ell\tau_{d_1}\cdots\tau_{d_n}\rangle_{g,n}
=\int_{\overline{\cal M}_{g,n}}
 \kappa^\ell\psi_1^{d_1}\cdots\psi_n^{d_n},\qquad
\ell,d_1,\ldots,d_n\in\bbZ_{\ge 0}
\end{align}
play the role of correlation functions.
They are defined on a closed Riemann surface $\Sigma$ of genus $g$
with $n$ marked points $p_1,\ldots,p_n$.
We let ${\cal M}_{g,n}$ denote the moduli space of $\Sigma$
and $\overline{\cal M}_{g,n}$
the Deligne-Mumford compactification of ${\cal M}_{g,n}$.
$\kappa$ (often denoted as $\kappa_1$ in the literature)
is the first Miller-Morita-Mumford
class, which is proportional to the Weil-Petersson symplectic form
$\omega=2\pi^2\kappa$.
$\psi_i$ is the first Chern class of the complex line bundle
whose fiber is the cotangent space to $p_i$
and $\tau_{d_i}=\psi_i^{d_i}$.
The intersection number in \eqref{eq:intersec} obeys the selection rule
\begin{align}
\int_{\overline{\cal M}_{g,n}}
 \kappa^\ell\psi_1^{d_1}\cdots\psi_n^{d_n}=0
\quad \mbox{unless}\quad \ell+d_1+\cdots+d_n=3g-3+n,
\label{eq:selection}
\end{align}
which we will use frequently.

For the above correlation functions
one can introduce the generating functions\footnote{
$G$ and $F$ in this paper are related to those in
our previous paper \cite{Okuyama:2019xbv}
by $G_{\rm here}=\gs^{-2}G_{\rm there}$,
$F_{\rm here}=\gs^{-2}F_{\rm there}$.}
\begin{align}
\label{eq:genGF}
\begin{aligned}
G(s,\{t_k\})&:=\sum_{g=0}^\infty \gs^{2g-2}G_g(s,\{t_k\}),&
F(\{t_k\})&:=\sum_{g=0}^\infty \gs^{2g-2}F_g(\{t_k\}),\\
G_g(s,\{t_k\})
 &:=\left\langle e^{s\kappa+\sum_{d=0}^\infty t_d\tau_d}\right\rangle_g,&
F_g(\{t_k\})
 &:=\left\langle e^{\sum_{d=0}^\infty t_d\tau_d}\right\rangle_g.
\end{aligned}
\end{align}
$G$ is actually expressed in terms of $F$ as
\cite{Mulase:2006baa,Dijkgraaf:2018vnm}
\begin{align}
\label{eq:GFrel}
G(s,\{t_k\})=F(\{t_k+\gamma_k s^{k-1}\}),
\end{align}
where $\ga_k$ is defined in \eqref{eq:gammavalue}.
Using this property
we showed \cite{Okuyama:2019xbv} that JT gravity is
nothing but the special case of
the general Witten-Kontsevich gravity with $t_k=\gamma_k$. 
Conversely, we can define a natural deformation of JT gravity
by (partially) releasing $t_k$ from the constraint $t_k=\ga_k$
and regard them as deformation parameters.
This is one of our main ideas in \cite{Okuyama:2019xbv}
and enables us to investigate JT gravity using the techniques
of the traditional 2d gravity.
In what follows we will apply this prescription
to multi-boundary correlators.

In this paper we study the $n$-boundary connected correlator
of JT gravity
\begin{align}
\label{eq:onshellcorr}
\langle Z(\beta_1)\cdots Z(\beta_n)\rangle_\conn.
\end{align}
We consider two kinds of its generalizations,
$Z_n(\beta_1,\ldots,\beta_n;t_0,t_1)$
and $Z_n(\beta_1,\ldots,\beta_n;\{t_k\})$.
They are obtained respectively by releasing only $t_0,t_1$
or all $\{t_k\}$ from the constraint $t_k=\gamma_k$.
We often call them ``off-shell'' correlators.
They are related to the ``on-shell'' correlators \eqref{eq:onshellcorr}
as
\begin{align}
\begin{aligned}
Z_n(\beta_1,\ldots,\beta_n,t_0,t_1,\{t_k=\gamma_k\}_{k\ge 2})
 &=Z_n(\beta_1,\ldots,\beta_n,t_0,t_1),\\
Z_n(\beta_1,\ldots,\beta_n,t_0=0,t_1=0)
 &=\langle Z(\beta_1)\cdots Z(\beta_n)\rangle_\conn.
\end{aligned}
\end{align}

As in \cite{Okuyama:2019xbv}
we introduce the notations
\begin{align}
\label{eq:rescaledvar}
\hbar:=\frac{1}{\sqrt{2}}\gs,\quad
x :=\hbar^{-1}t_0,\quad
\tau :=\hbar^{-1}t_1
\end{align}
and
\begin{align}
\label{eq:rescaleddiff}
\partial_k :=\frac{\partial}{\partial t_k},\quad
{}^{'}:=\partial_x=\hbar\partial_0,\quad
\dot{~}:=\partial_\tau =\hbar\partial_1.
\end{align}
As discussed in \cite{Okuyama:2019xbv},
$\gs$ is the genus-counting parameter in the high temperature regime while
$\hbar$ is the natural coupling constant in the low temperature 't Hooft limit 
\eqref{eq:tHooft}.
It is also convenient to introduce
\begin{align}
\label{eq:defIn}
I_n(v,\{t_k\})=\sum_{\ell=0}^\infty t_{n+\ell}\frac{v^\ell}{\ell!}
\quad (n\ge 0),
\end{align}
because it is known \cite{Itzykson:1992ya} that
$F_g(\{t_k\})\ (g\ge 2)$ are polynomials
in $I_n(u_0,\{t_k\})\ (n\ge 2)$
and in $[1-I_1(u_0,\{t_k\})]^{-1}$,
with
\begin{align}
\label{eq:defu0}
u_0&:=\partial_0^2 F_0.
\end{align}

We will work mostly on the partially constrained case
$t_k=\gamma_k\ (k\ge 2)$, leaving $t_0$ and $t_1$ as parameters.
In this case
we further introduce the new variables
\begin{align}
\label{eq:yt-def}
y:=u_0,\quad
t:=1-I_1
\end{align}
and the functions
\begin{align}
\label{eq:Bndef}
B_n(y)
 :=\frac{J_n(2\sqrt{y})}{y^{n/2}}
  =\sum_{k=\max(0,-n)}^\infty\frac{(-1)^ky^k}{k!(k+n)!}\qquad(n\in\bbZ).
\end{align}
Here, $J_n(z)$ is the Bessel function of the first kind.
$B_n$ are related to $I_n$ as
\begin{align}
B_{n-1}&=(-1)^n I_n(y,\{t_0,t_1,t_k=\gamma_k\ (k\ge 2)\})
\quad (n\ge 2)
\end{align}
and satisfy
\begin{align}
\partial_yB_n&=-B_{n+1},\qquad
yB_{n+1}=nB_n-B_{n-1}.
\end{align}
The old variables $(t_0,t_1)$ and the new ones $(y,t)$
are related as
\begin{align}
t_1=B_0-t,\quad
t_0=y(B_1-t_1).
\label{eq:t1t0}
\end{align}
The differentials $\partial_{0,1}$
are then written in the new variables as\footnote{
This change of variables was originally introduced by Zograf
(see e.g.~\cite{Zograf:2008wbe}).}
\begin{align}
\partial_0=\frac{1}{t}(\partial_y-B_1\partial_t),\quad
\partial_1=y\partial_0-\partial_t.
\end{align}
The on-shell value 
$(t_0,t_1)=(0,0)$ corresponds to $(y,t)=(0,1)$.
At this value $B_n$ becomes
\begin{align}
\label{eq:Bnonshell}
B_n(0)=\frac{1}{n!},\quad n\ge 0.
\end{align}
%

\subsection{Multi-boundary correlators of general topological gravity
            \label{sec:gexpMulti}}

In our previous paper \cite{Okuyama:2019xbv}
we formulated how to compute the genus expansion of
the partition function $Z_1=Z_\JT$ of JT gravity on a surface
with one boundary.
In this section we generalize the method
to the case of multi-boundary correlators of general topological
gravity.

Let us start with the fact that the $n$-boundary correlator
of general topological gravity is given by \cite{Moore:1991ir}
\begin{align}
\label{eq:ZninF}
\begin{aligned}
Z_n(\{\beta_i\},\{t_k\})
 &\simeq B(\beta_1)\cdots B(\beta_n)F(\{t_k\}).
\end{aligned}
\end{align}
Here $F$ is
defined in \eqref{eq:genGF} and the operator $B(\beta)$ is given by
\begin{align}
\label{eq:Bdef}
B(\beta)
 =\gs\sqrt{\frac{\beta}{2\pi}}\sum_{d=0}^\infty\beta^d\partial_d.
\end{align}
As mentioned in section \ref{sec:intro},
\eqref{eq:Bdef} is based on the idea that
the macroscopic loop operator $Z(\bt)$ is expanded
in terms of the microscopic loop operator
$\tau_d$ in the limit $\bt\to0$
\begin{equation}
\begin{aligned}
 Z(\bt)\simeq \gs\sqrt{\frac{\beta}{2\pi}}\sum_{d=0}^\infty\beta^d\tau_d
\end{aligned} 
\end{equation}
and the insertion of $\tau_d$ is represented by the derivative
$\del_d$ when acting on the free energy $F$.
$B(\beta)$ can be thought of as the ``boundary creation operator.''
We put the symbol ``$\simeq$'' in \eqref{eq:ZninF}, meaning that
the equality holds up to
an additional non-universal part \cite{Moore:1991ir}
when $3g-3+n<0$. Such a deviation appears, however, only
in the genus-zero part of $n=1,2$-boundary correlators,
which we will discuss separately in section~\ref{sec:genus0}.
Note that
the complex dimension of the moduli space $\b{\mathcal{M}}_{g,n}$ is
$3g-3+n$ which becomes negative for $(g,n)=(0,1)$ and $(0,2)$.

As in the case of single boundary 
the genus expansion can be computed by solving a differential
equation which follows from the KdV equation.
To see this, let us first introduce
\begin{align}
\begin{aligned}
W_n(\{\beta_i\},\{t_k\})
 :=&\ \partial_x Z_n = \hbar\partial_0 Z_n\\
  \simeq&\ \hbar B(\beta_1)\cdots B(\beta_n)\partial_0 F,\\
W_0(\{t_k\}):=&\ \partial_x F = \hbar\partial_0 F,\\
u:=&\ \gs^2\partial_0^2 F=2\del_x^2F.
\end{aligned}
\label{eq:Wudef}
\end{align}
Recall that $u$ satisfies the KdV equation
\cite{Witten:1990hr,Kontsevich:1992ti}
\begin{align}
\label{eq:uKdV}
\dot{u}=uu'+\frac{1}{6}u'''.
\end{align}
Integrating this equation once in $x=\hbar^{-1}t_0$ we have
\begin{align}
\label{eq:W0KdV}
\dot{W}_0
 &=(W_0')^2+\frac{1}{6}W_0'''.
\end{align}
Since $B(\beta_i)$ commutes with
$\dot{~}=\partial_\tau$ and ${}'=\partial_x$,
we immediately obtain a differential equation for $W_n$
by simply acting $B(\beta_1)\cdots B(\beta_n)$
on both sides of the above equation.
For instance, by acting $B(\beta_1)$ on both sides of 
\eqref{eq:W0KdV} we obtain
\begin{align}
\label{eq:W1KdV}
\begin{aligned}
\dot{W}_1
 &=2W_0'W_1'+\frac{1}{6}W_1'''\\
 &=uW_1'+\frac{1}{6}W_1'''
\end{aligned}
\end{align}
for $W_1(\beta_1)$.
This is nothing but the differential equation for $W$
in \cite{Okuyama:2019xbv} with the identification $W_1=W$.
By further acting $B(\beta_2)$ on both sides of \eqref{eq:W1KdV}
we obtain
\begin{align}
\label{eq:W2KdV}
\begin{aligned}
\dot{W}_2(\beta_1,\beta_2)
 &=2W_1'(\beta_1)W_1'(\beta_2)+2W_0'W_2'(\beta_1,\beta_2)
  +\frac{1}{6}W_2'''(\beta_1,\beta_2)\\
 &=2W_1'(\beta_1)W_1'(\beta_2)+uW_2'(\beta_1,\beta_2)
  +\frac{1}{6}W_2'''(\beta_1,\beta_2).
\end{aligned}
\end{align}
In general the differential equation for $W_n$ may be written as
\begin{align}
\label{eq:WnKdV}
\begin{aligned}
\dot{W}_n(\beta_1,\ldots,\beta_n)
 &=\sum_{I\subset N}W_{|I|}'W_{|N-I|}'
   +\frac{1}{6}W_n'''(\beta_1,\ldots,\beta_n),
\end{aligned}
\end{align}
where $N=\{1,2,\ldots,n\}$,
$W_{|I|}'=W_{|I|}'(\beta_{i_1},\ldots,\beta_{i_{|I|}})$
with $I=\{i_1,i_2,\ldots,i_{|I|}\}$
and the sum is taken for all possible subset $I$ of $N$
including the empty set.
The equation \eqref{eq:WnKdV} uniquely determines
$W_n$ in the genus expansion
given the genus expansion of $W_k\ (k<n)$
and the genus zero part $W_n^{g=0}$.
It is important to note that the non-universal parts are
entirely absent
in \eqref{eq:WnKdV}
because all the elements other than $W_0'=\frac{u}{2}$
appearing in \eqref{eq:WnKdV}
are equal to or higher than the third derivative of $F_0$
(see the discussion in the next subsection).

Finally $Z_n$ is obtained by merely integrating $W_n$ once
in $x$.
This can be done order by order in the genus expansion.
As a demonstration
we will study in detail
the case of two-boundary correlator of JT gravity
in subsection~\ref{sec:gexp2pt}.

\subsection{Genus zero part\label{sec:genus0}}

In this section let us consider the genus zero part of
the multi-boundary correlator $Z_n^{g=0}$ and calculate
$W_n^{g=0}=\partial_x Z_n^{g=0}$.
In fact, $Z_n^{g=0}$ has been already calculated in the literature
\cite{Ambjorn:1990ji,Moore:1991ir,Ginsparg:1993is}.
In what follows we will reproduce the results in our notation.

Restricting \eqref{eq:ZninF} to the genus zero part, we have
\begin{align}
\label{eq:Zng0inF}
Z_n^{g=0}(\{\beta_i\})\simeq\gs^{-2}B(\beta_1)\cdots B(\beta_n)F_0.
\end{align}
Recall that $F_0$ is expressed as \cite{Itzykson:1992ya}
\begin{align}
\label{eq:F0IZ}
F_0
 &=\frac{1}{2}\int_0^{u_0}dv\left(I_0(v,\{t_k\})-v\right)^2,
\end{align}
where $I_0$ and $u_0$ are defined in \eqref{eq:defIn}--\eqref{eq:defu0}.
Note that for $v=u_0$ we have
\begin{align}
u_0=I_0(u_0,\{t_k\}).
\end{align}
Using these relations we obtain
\begin{align}
\label{eq:Z1g0}
\begin{aligned}
Z_1^{g=0}(\beta,\{t_k\})
 &\simeq\gs^{-2}B(\beta)F_0\\
 &=\frac{1}{\gs}\sqrt{\frac{\beta}{2\pi}}
   \sum_{d=0}^\infty\beta^d
   \int_0^{u_0}dv\left(I_0(v,\{t_k\})-v\right)\partial_d I_0(v,\{t_k\})\\
 &=\frac{1}{\gs}\sqrt{\frac{\beta}{2\pi}}
   \int_0^{u_0}dv\left(I_0(v,\{t_k\})-v\right)e^{\beta v}.
\end{aligned}
\end{align}
As mentioned above, the last expression is only reliable up to
the non-universal part.
The non-universal part arises because the correlator is not
fully constrained by the intersection number of quantum gravity 
which is defined only for $3g-3+n\ge 0$. 
However, by taking derivative with respect to $t_k$
we can insert the microscopic loop operator 
$\tau_k$ into the bracket of the intersection number
and increase $n$ by one.
By repeating this procedure we can map the computation of
the partition function precisely to the integral over
the well-defined moduli space of punctured Riemann surfaces.
For the one-point function
we can remove the non-universal part by differentiating
twice in $t_0$
\begin{align}
\label{eq:ddZ1g0}
\begin{aligned}
\partial_0^2Z_1^{g=0}(\beta,\{t_k\})
 &=\frac{1}{\gs}\sqrt{\frac{\beta}{2\pi}}
   \partial_0^2\int_0^{u_0}dv\left(I_0(v,\{t_k\})-v\right)e^{\beta v}\\
 &=\frac{1}{\gs}\sqrt{\frac{\beta}{2\pi}}
   \partial_0\int_0^{u_0}dv e^{\beta v}\\
 &=\frac{1}{\gs}\sqrt{\frac{\beta}{2\pi}}
   \partial_0\frac{e^{\beta u_0}}{\beta}.
\end{aligned}
\end{align}

$Z_1^{g=0}$ is obtained by integrating the above relation
twice in $t_0$.
We impose the boundary condition that $Z_1^{g=0}$
identically vanishes
for $u_0\to -\infty$
\begin{align}
\label{eq:Z1g0bc}
Z_1^{g=0}\big|_{u_0=-\infty}=\partial_0 Z_1^{g=0}\big|_{u_0=-\infty}=0.
\end{align}
This is naturally understood from our viewpoint \cite{Okuyama:2019xbv}
that $Z_1=Z_\JT=\Tr(e^{\beta Q}\Pi)$ is the macroscopic loop operator,
in which $Q=\partial_x^2+u$ is approximated as $Q\sim u_0$ at genus zero.
Hence we have
\begin{align}
\label{eq:Z1g0true}
\begin{aligned}
Z_1^{g=0}(\beta,\{t_k\})
 &=\frac{1}{\gs}\sqrt{\frac{\beta}{2\pi}}
   \int_{-\infty}^{u_0}dv\left(I_0(v,\{t_k\})-v\right)e^{\beta v}.
\end{aligned}
\end{align}
In other words the true genus-zero part of the one-point function
\eqref{eq:Z1g0true} including the non-universal term
is obtained from \eqref{eq:Z1g0} by replacing the
region of integration from $[0,u_0]$ to $(-\infty,u_0]$.
Note that if we set $t_k=\gamma_k\ (k\ge 2)$
the above expression gives the result for JT gravity
\begin{align}
Z_1^{g=0}(\beta,t_0,t_1)
 =\frac{1}{2\sqrt{\pi\beta}\hbar}
  \int_{-\infty}^{u_0}dv\left(J_0(2\sqrt{v})-t_1\right)e^{\beta v}.
\end{align} 
By further setting $t_1=0$ this reproduces
our previous result obtained in \cite{Okuyama:2019xbv}.

In a similar manner, we can compute the
genus zero part of the two-boundary correlator
\begin{align}
\label{eq:Z2g0}
\begin{aligned}
Z_2^{g=0}(\beta_1,\beta_2,\{t_k\})
 &\simeq\gs^{-2}B(\beta_1)B(\beta_2)F_0\\
 &=\sqrt{\frac{\beta_1\beta_2}{(2\pi)^2}}
   \sum_{d=0}^\infty\beta_1^d\partial_d
   \int_0^{u_0}dv\left(I_0(v,\{t_k\})-v\right)e^{\beta_2 v}\\
 &=\sqrt{\frac{\beta_1\beta_2}{(2\pi)^2}}
   \int_0^{u_0}dv e^{(\beta_1+\beta_2)v}\\
 &=\sqrt{\frac{\beta_1\beta_2}{(2\pi)^2}}
   \frac{e^{(\beta_1+\beta_2)u_0}-1}{\beta_1+\beta_2}.
\end{aligned}
\end{align}
We can remove the non-universal part by differentiating
once in $t_0$
\begin{align}
\label{eq:dZ2g0}
\begin{aligned}
\partial_0 Z_2^{g=0}(\beta_1,\beta_2,\{t_k\})
 &=\sqrt{\frac{\beta_1\beta_2}{(2\pi)^2}}
   \partial_0\frac{e^{(\beta_1+\beta_2)u_0}}{\beta_1+\beta_2}.
\end{aligned}
\end{align}
By imposing the boundary condition
\begin{align}
Z_2^{g=0}\big|_{u_0=-\infty}=0
\end{align}
we obtain
\begin{align}
\label{eq:Z2g0true}
\begin{aligned}
Z_2^{g=0}(\beta_1,\beta_2,\{t_k\})
 &=\sqrt{\frac{\beta_1\beta_2}{(2\pi)^2}}
   \frac{e^{(\beta_1+\beta_2)u_0}}{\beta_1+\beta_2}.
\end{aligned}
\end{align}
Again the true two-point function is obtained from \eqref{eq:Z2g0}
by extending the integration region to $(-\infty,u_0]$.
Given this expression we can easily
determine the genus-zero part of the $n$-point function
by induction in $n$
\begin{align}
\label{eq:Zng0main}
\begin{aligned}
Z_n^{g=0}(\{\beta_i\},\{t_k\})
 &=\sqrt{\frac{\prod_{i=1}^n\beta_i}{(2\pi)^n}}
   \frac{(\gs\partial_0)^{n-2}e^{\sum_{i=1}^n\beta_i u_0}}
        {\sum_{i=1}^n\beta_i}\quad (n\ge 2),
\end{aligned}
\end{align}
where we have used the genus-zero version of the KdV flow\footnote{
\eqref{eq:zero-flow} can also be shown by using 
$\partial_k u_0=\del_k\del_0^2F_0$ with $F_0$ in \eqref{eq:F0IZ}. 
}
\begin{align}
\partial_k u_0=\partial_0\cR_{k+1}=\partial_0\frac{u_0^{k+1}}{(k+1)!}.
\label{eq:zero-flow}
\end{align}

Finally, the genus zero part $W_n^{g=0}=\partial_x Z_n^{g=0}$
is obtained
from \eqref{eq:ddZ1g0}, \eqref{eq:Z1g0bc} and \eqref{eq:Zng0main} as
\begin{align}
\label{eq:Wng0}
\begin{aligned}
W_n^{g=0}(\{\beta_i\},\{t_k\})
 &=\sqrt{\frac{\prod_{i=1}^n\beta_i}{2(2\pi)^n}}
   \frac{(\gs\partial_0)^{n-1}e^{\sum_{i=1}^n\beta_i u_0}}
        {\sum_{i=1}^n\beta_i}\quad (n\ge 1).
\end{aligned}
\end{align}
%

\subsection{Two-boundary correlator of JT gravity\label{sec:gexp2pt}}

In this section we focus on the two-boundary correlator of JT gravity
and demonstrate in detail how to compute the genus expansion
by the method described in section~\ref{sec:gexpMulti}.

Before explaining our method,
let us first briefly recall how to compute the correlator
by using the method of \cite{Saad:2019lba}.
The correlator is evaluated by the path-integral of JT gravity
on two-dimensional surfaces of arbitrary genus with two boundaries.
As shown in \cite{Saad:2019lba}, the $n$-boundary
correlator is written as 
a combination of simple building blocks:
the partition function of Schwarzian mode on the ``trumpet geometry''
$Z_{\text{trumpet}}(\bt_i,b_i)$ and the Weil-Petersson volume 
$V_{g,n}(b_1,\cdots,b_n)$ of the moduli space of
Riemann surface with geodesic boundaries with lengths
$b_i\ (i=1,\cdots,n)$
\begin{equation}
\begin{aligned}
Z_{\text{trumpet}}(\bt,b)
 &=\frac{e^{-\frac{\gamma b^2}{2\beta}}}{\sqrt{2\pi\beta\gamma^{-1}}},\\
V_{g,n}(b_1,\cdots,b_n)
 &=\left\langle
   \exp\biggl(2\pi^2\kappa+\sum_{i=1}^n\frac{b_i^2}{2}\psi_i\biggr)
   \right\rangle_{g,n},
\end{aligned} 
\label{eq:trumpet}
\end{equation}
where $\ga$ is the asymptotic value of the dilaton field
at the boundary of spacetime.
Then the genus sum of two-boundary correlator is written as
\begin{align}
\label{eq:Z2decomp}
\begin{aligned}
\langle Z(\beta_1)Z(\beta_2)\rangle_\conn
 &=\langle Z(\beta_1)Z(\beta_2)\rangle_\conn^{g=0}
  +\langle Z(\beta_1)Z(\beta_2)\rangle_\conn^{g\ge 1},
\end{aligned}
\end{align}
where $g=0$ and $g\ge 1$ parts are evaluated respectively as
\begin{align}
\label{eq:gexpZ2formal}
\begin{aligned}
\langle Z(\beta_1)Z(\beta_2)\rangle_\conn^{g=0}
 &=\int_0^\infty bdb
   Z_{\text{trumpet}}(\bt_1,b) Z_{\text{trumpet}}(\bt_2,b)
  =\frac{\sqrt{\beta_1\beta_2}}{2\pi(\beta_1+\beta_2)},\\
\langle Z(\beta_1)Z(\beta_2)\rangle_\conn^{g\ge 1}
 &=\sum_{g=1}^\infty e^{-2gS_0}
  \int_0^\infty
  \prod_{i=1,2}b_idb_i  Z_{\text{trumpet}}(\bt_i,b_i)
  V_{g,2}(b_1,b_2)\\
 &=\sum_{g=1}^\infty e^{-2gS_0}
  \frac{\sqrt{\beta_1\beta_2\gamma^{-2}}}{2\pi}
  \left\langle
   \frac{e^{2\pi^2\kappa}}{\prod_{i=1,2}(1-\beta_i\gamma^{-1}\psi_i)}
  \right\rangle_{g,2}\\
 &=\frac{\sqrt{\beta_1\beta_2}}{2\pi}
  \sum_{g=1}^\infty \gs^{2g}
  \left\langle
   \frac{e^{\kappa}}{\prod_{i=1,2}(1-\beta_i\psi_i)}
  \right\rangle_{g,2}.
\end{aligned}
\end{align}
Here we have set 
\begin{equation}
\begin{aligned}
\gamma=\frac{1}{2\pi^2},\qquad
 \gs=(2\pi^2)^{\frac{3}{2}}e^{-S_0}
\end{aligned} 
\label{eq:ga-gs}
\end{equation}
as in \cite{Okuyama:2019xbv}
and we have used the selection rule \eqref{eq:selection}.
From the above expressions one obtains
\begin{align}
\label{eq:gexpZ2result}
\langle Z(\beta_1)Z(\beta_2)\rangle_\conn
=\frac{\sqrt{\beta_1\beta_2}}{2\pi}
\left[\frac{1}{\beta_1+\beta_2}
 +\left(\frac{1}{16}+\frac{\beta_1+\beta_2}{12}
       +\frac{\beta_1^2+\beta_1\beta_2+\beta_2^2}{24}\right)\gs^2
 +{\cal O}(\gs^4)\right].
\end{align}
This expansion can be computed up to arbitrary genus in principle
given the data of
$\langle\kappa^\ell\psi_1^{d_1}\psi_2^{d_2}\rangle_{g,2}$.

Let us now move on to explaining
our method described in section~\ref{sec:gexpMulti}.
Using this method the genus expansion \eqref{eq:gexpZ2result}
can be computed very efficiently,
as in the case of one-boundary partition function \cite{Okuyama:2019xbv}.
Regarding the genus zero result \eqref{eq:Wng0}
we first expand $u,W_1,W_2$ as\footnote{
$W_{g,1}$ is related to $W_g$ in \cite{Okuyama:2019xbv}
by $\beta W_{g,1}=W_g$.}
\begin{align}
\label{eq:uW1W2exp}
\begin{aligned}
u&=\sum_{g=0}^\infty \gs^{2g}u_g,\\
W_1(\beta)
 &=\sqrt{\frac{\beta}{4\pi}}e^{\beta y}
   \sum_{g=0}^\infty\gs^{2g}W_{g,1}(\beta),\\
W_2(\beta_1,\beta_2)
 &=\gs\sqrt{\frac{\beta_1\beta_2}{8\pi^2}}e^{(\beta_1+\beta_2)y}
   \sum_{g=0}^\infty\gs^{2g}W_{g,2}(\beta_1,\beta_2).
\end{aligned}
\end{align}
The genus zero coefficients are given respectively as
\begin{align}
\label{eq:uW1W2init}
u_0=y,\qquad
W_{0,1}(\beta)=\frac{1}{\beta},\qquad
W_{0,2}(\beta_1,\beta_2)=\frac{1}{t}.
\end{align}
By plugging the expansions \eqref{eq:uW1W2exp}
into the differential equations
\eqref{eq:uKdV}, \eqref{eq:W1KdV} and \eqref{eq:W2KdV}
we obtain the recursion relations
\begin{align}
\label{eq:uW1W2recrel}
\begin{aligned}
-\frac{1}{t}\partial_t(tu_g)
 &=\sum_{h=1}^g u_{g-h}\partial_0 u_h
   +\frac{1}{12}\partial_0^3 u_{g-1},\\
-\partial_t W_{g,1}(\beta)
 &=\sum_{h=1}^g u_h\partial_{0,\beta}W_{g-h,1}(\beta)
 +\frac{1}{12}\partial_{0,\beta}^3 W_{g-1,1}(\beta),\\
-\partial_t W_{g,2}(\beta_1,\beta_2)
 &=\sum_{h=0}^g\partial_{0,\beta_1}W_{h,1}(\beta_1)
               \partial_{0,\beta_2}W_{g-h,1}(\beta_2)\\
 &\hspace{1em}
 +\sum_{h=1}^g u_h\partial_{0,\beta_1+\beta_2}W_{g-h,2}(\beta_1,\beta_2)
 +\frac{1}{12}\partial_{0,\beta_1+\beta_2}^3W_{g-1,2}(\beta_1,\beta_2),
\end{aligned}
\end{align}
where we have introduced the notation
\begin{align}
\partial_{0,\beta}
 :=e^{-\beta y}\partial_0 e^{\beta y}
 =\partial_0+\beta t^{-1}.
\end{align}
The higher genus coefficients $u_g,W_{g,1},W_{g,2}$
are computed by solving these recursion relations
with the initial data \eqref{eq:uW1W2init}.

We next expand $Z_2$ as
\begin{align}
\label{eq:Z2gexp}
\begin{aligned}
Z_2(\beta_1,\beta_2)
 &=\frac{\sqrt{\beta_1\beta_2}}{2\pi}
  e^{(\beta_1+\beta_2)y}
  \sum_{g=0}^\infty\gs^{2g}Z_{g,2}(\beta_1,\beta_2).
\end{aligned}
\end{align}
The coefficient $Z_{g,2}$ is then obtained from the relation
\begin{align}
\partial_{0,\beta_1+\beta_2}Z_{g,2}(\beta_1,\beta_2)
 =W_{g,2}(\beta_1,\beta_2).
\end{align}
As in \cite{Okuyama:2019xbv},
the integration in $t_0$ can be done unambiguously
assuming that $Z_{g,2}\ (g\ge 1)$ is a polynomial in $t^{-1}$
without $t$-independent term. We find
\begin{align}
\label{eq:Zg2results}
Z_{0,2}=\frac{1}{\beta_1+\beta_2},\quad
Z_{1,2}=
  \frac{\beta_1^2+\beta_1\beta_2+\beta_2^2}{24t^2}
 +\frac{2(\beta_1+\beta_2)B_1-B_2}{24t^3}
 +\frac{B_1^2}{12t^4},\quad\cdots.
\end{align}
Setting $(y,t)=(0,1)$
with the on-shell values \eqref{eq:Bnonshell} of $B_n$
one can check that \eqref{eq:Z2gexp} with \eqref{eq:Zg2results}
reproduces the expansion \eqref{eq:gexpZ2result}.

\subsection{On multi-boundary correlator of JT gravity}

Using the method of \cite{Saad:2019lba}
the $n$-boundary connected correlator for $n\ge 3$ is
obtained by combining the contribution of $n$ trumpets 
and the Weil-Petersson volume in \eqref{eq:trumpet}
\begin{align}
\begin{aligned}
&\hspace{-1em}
\langle Z(\beta_1)\cdots Z(\beta_n)\rangle_\conn\\
 &=\sum_{g=0}^\infty e^{-(2g-2+n)S_0}
  \int_0^\infty\prod_{i=1}^nb_idb_i Z_{\text{trumpet}}(\bt_i,b_i)
  V_{g,n}(b_1,\cdots,b_n)\\
 &=\sum_{g=0}^\infty e^{-(2g-2+n)S_0}
  \sqrt{\frac{\prod_{i=1}^n\beta_i}{(2\pi\gamma)^{n}}}
  \left\langle
   \frac{e^{2\pi^2\kappa}}{\prod_{i=1}^n(1-\beta_i\gamma^{-1}\psi_i)}
  \right\rangle_{g,n}\\
 &=\sqrt{\frac{\prod_{i=1}^n\beta_i}{(2\pi)^{n}}}
  \sum_{g=0}^\infty \gs^{2g-2+n}
  \left\langle
   \frac{e^{\kappa}}{\prod_{i=1}^n(1-\beta_i\psi_i)}
  \right\rangle_{g,n},
\end{aligned}
\end{align}
where we have set $\ga$ and $\gs$ as in \eqref{eq:ga-gs}
and have used the selection rule \eqref{eq:selection}.
This expression is reproduced from \eqref{eq:ZninF} as follows.
For $n\ge 3$ we have
\begin{align}
\begin{aligned}
 \langle Z(\beta_1)\cdots Z(\beta_n)\rangle_\conn
 &=Z_n(\{\beta_i\},\{t_k=\gamma_k\})\\
 &=B_1(\beta_1)\cdots B_n(\beta_n)F\Big|_{t_k=\gamma_k}.
\end{aligned}
\end{align}
Note that the non-universal part is absent for $n\geq3$ since
the dimension of the moduli space $3g-3+n$ is non-negative in this case.
Using the relation \eqref{eq:GFrel} between $F$ and $G$ we have
\begin{align}
\begin{aligned}
 \langle Z(\beta_1)\cdots Z(\beta_n)\rangle_\conn
 &=B_1(\beta_1)\cdots B_n(\beta_n)G\big|_{s=1,t_k=0}\\
 &=\sqrt{\frac{\prod_{i=1}^n\beta_i}{(2\pi)^{n}}}
 \sum_{g=0}^\infty\gs^{2g-2+n}
 \left\langle\frac{e^{\kappa}}{\prod_{i=1}^n(1-\beta_i\psi_i)}
 \right\rangle_{g,n},
\end{aligned}
\end{align}
where we have used
$\tau_{d_i}=\psi_i^{d_i}$ as in \cite{Okuyama:2019xbv}.

\section{WKB and 't Hooft expansions}\label{sec:wkb}

In this section we study the WKB expansion of the Baker-Akhiezer
wave function and the 't Hooft expansion of
the multi-boundary correlators.
Our new formalism is much more efficient than our previous method
based on the topological recursion \cite{Okuyama:2019xbv}.

\subsection{Baker-Akhiezer function\label{sec:BAgexp}}

As we saw in \cite{Okuyama:2019xbv}
the Baker-Akhiezer functions $\psi_\pm(\xi;\{t_k\})$ are
certain two independent solutions of the Schr\"odinger equation
\begin{align}
\label{eq:Schrxi}
Q\psi&=\xi\psi,
\end{align}
where
\begin{align}
Q:=\partial_x^2+u
\end{align}
and $u$ is defined in \eqref{eq:Wudef}.
More specifically, in terms of the resolvent
\begin{align}
R(\xi)
 =\Bigl\langle x\,\Big|\,\frac{1}{\xi-Q}\,\Big|\,x\Bigr\rangle
 =\int_0^\infty d\beta e^{-\beta\xi}W_1(\beta),
\end{align}
$\psi_\pm$ are expressed as
\begin{align}
\psi_\pm = \sqrt{R}e^{\pm S},\quad
S'=\frac{1}{2R}.
\end{align}

Let us introduce
\begin{align}
\begin{aligned}
A_\pm
:=&\,\log\psi_\pm
 =\pm S +\frac{1}{2}\log R,\\
v_\pm
:=&\,A_\pm' 
 =\frac{\pm 1+R'}{2R}.
\end{aligned}
\end{align}
From the differential equation \cite{Gelfand:1975rn,BBT}
\begin{align}
2RR''-{R'}^2+4(u-\xi)R^2=-1
\end{align}
we see that
$v_\pm$ are solutions to the equation
\begin{align}
\label{eq:Burel}
v^2+v'=\xi-u.
\end{align}
Using this equation we can compute the WKB expansion
of $v_\pm$ as follows.
Let us assume that $v$ admits the expansion
\begin{align}
v=\sum_{n=0}^\infty\hbar^nv_n.
\end{align}
By plugging this form as well as the 
genus expansion of $u$
\begin{align}
u=\sum_{g=0}^\infty(\sqrt{2}\hbar)^{2g}u_g
\end{align}
into \eqref{eq:Burel}, we find
\begin{align}
v_0^2=\xi-u_0,\quad 2v_0v_1+\partial_0v_0=0
\end{align}
at the leading and the next to the leading orders.
From these we find\footnote{
For the sake of simplicity we restrict ourselves
hereafter to the JT gravity case $t_k=\gamma_k\ (k\ge 2)$,
i.e.~we set $1-I_1=t$,
but the discussion here 
would be easily generalized to
the case of general topological gravity.}
\begin{align}
\label{eq:Binit}
v_0=\pm z,\quad
v_1
 =-\frac{1}{2}\partial_0\log v_0
 =\frac{\partial_0 u_0}{4(\xi-u_0)}
 =\frac{1}{4tz^2},
\end{align}
where we have introduced the notation
\begin{align}
z := \sqrt{\xi-u_0}.
\end{align}
At the order of $\hbar^n\ (n\ge 2)$
\eqref{eq:Burel} is written as the recursion relation
\begin{align}
v_n
 =-\frac{1}{2v_0}
   \left(\partial_0 v_{n-1}+\sum_{k=1}^{n-1}v_k v_{n-k}
        +\left\{\begin{array}{ll}
2^{\frac{n}{2}}u_{\frac{n}{2}}&\mbox{($n$ even)}\\
                           0&\mbox{($n$ odd)}
		 \end{array}\right.
   \right).
\end{align}
Solving this recursion relation
with the initial condition $v_0=+z$
one can easily calculate the WKB expansion of $v_+$.
(Recall that the genus expansion of $u$ is calculated by
solving \eqref{eq:uW1W2recrel}.)
In the same way one can calculate $v_-$ starting from
the initial condition $v_0=-z$, but instead
$v_-$ is obtained from $v_+$ by merely replacing $z$ with $-z$.
The same arguments hold for $A_\pm$ and $\psi_\pm$.
Therefore, in what follows we omit the subscript ``$+$'' and write
\begin{align}
v_+ =v,\quad
A_+&=A,\quad
\psi_+=\psi,
\end{align}
with the understanding that
\begin{align}
v_- =v|_{z\to -z},\quad
A_- =A|_{z\to -z},\quad
\psi_- =\psi|_{z\to -z}.
\end{align}
As a side remark, note that
\eqref{eq:Burel} is viewed as a Miura transformation.
From this viewpoint $v$ can be viewed as a solution
to the modified KdV equation
\begin{align}
\label{eq:mKdV}
\tpartial_1 v=\frac{\hbar^2}{6}\partial_0^3 v-v^2\partial_0 v
\end{align}
with
\begin{align}
\tpartial_1:=\partial_1-\xi\partial_0.
\end{align}
It is also possible to compute the WKB expansion of $v$
by directly solving this equation
with the initial condition \eqref{eq:Binit}.

Using the above method we obtain
\begin{align}
v_0=z,\quad
v_1=\frac{1}{4tz^2},\quad
v_2=-\frac{5}{32t^2z^5}
 +\frac{1}{t^3}\left(-\frac{B_1}{8z^3}+\frac{B_2}{24z}\right)
 -\frac{B_1^2}{12t^4z},\quad\cdots.
\end{align}
Next, let us consider the WKB expansion of $A=\log\psi$.
We expand $A$ as
\begin{align}
A=\sum_{n=0}^\infty\hbar^{n-1}A_n,
\end{align}
so that we have
\begin{align}
\label{eq:diffAn}
\partial_0A_n=v_n.
\end{align}
By solving this we find
\begin{align}
\label{eq:An}
\begin{aligned}
A_0
 &=-\frac{2}{3}tz^3
   +\sum_{n=1}^\infty\frac{(-2)^{n+1}B_n}{(2n+3)!!}z^{2n+3},\\
A_1&=-\frac{1}{2}\log|z|-\hf\log(4\pi),\\
A_2&=-\frac{5}{48tz^3}-\frac{B_1}{24t^2z},\\
A_3&=\frac{5}{64t^2z^6}
     +\frac{1}{t^3}\left(\frac{B_1}{16z^4}-\frac{B_2}{48z^2}\right)
     +\frac{B_1^2}{24t^4z^2}.
\end{aligned}
\end{align}
Here 
$A_1$ is immediately obtained from \eqref{eq:Binit}
up to the integration constant $-\hf\log(4\pi)$. This
constant is universal in the sense that it does not depend on the background.
Thus it can be determined by the asymptotic expansion of the
Airy function which is the BA function for the pure topological gravity
corresponding to the trivial background
$t_n=0~(n\geq1)$.\footnote{As reviewed in
appendix A of \cite{Okuyama:2019xbv}, 
the BA function for the pure topological gravity is given by
$\psi=\hbar^{-\frac{1}{6}}\text{Ai}\bigl(\hbar^{-\frac{2}{3}}z^2\bigr)$
which has the  large $z$ asymptotic expansion
$
 \psi\approx \frac{1}{\rt{4\pi z}}e^{-\frac{2z^3}{3\hbar}}
$.}
$A_n\ (n\ge 2)$ is also easily obtained 
assuming that $A_n$ is a polynomial in $t^{-1}$
without $t$-independent term.
Getting $A_0$ is less trivial, but
one can explicitly check that $A_0$ given in \eqref{eq:An}
satisfies \eqref{eq:diffAn}.
One can also check that
\begin{align}
V_\eff(\xi)
 \equiv-2A_0
 =\frac{4}{3}tz^3
  +\sum_{n=1}^\infty\frac{(-1)^n(n+1)!B_n}{(2n+3)!}(2z)^{2n+3}
\label{eq:off-Veff}
\end{align}
is regarded as the off-shell generalization of
the effective potential $V_\eff(\xi)$
discussed in \cite{Saad:2019lba,Okuyama:2019xbv}:
\begin{align}
V_\eff(\xi)\Big|_{y=0,t=1}
 =\frac{1}{2}\sin\bigl(2\sqrt{\xi}\bigr)
  -\sqrt{\xi}\cos\bigl(2\sqrt{\xi}\bigr).
\end{align}
In this way one can in principle compute the WKB expansion
of $\psi=\exp\left[\sum_{n=0}^\infty\hbar^{n-1}A_n\right]$
up to any order.

We note in passing that
the on-shell BA function and its derivative are expanded as
\begin{equation}
\begin{aligned}
 \psi&=\sum_{n=0}^\infty \hbar^{\frac{2n}{3}-\frac{1}{6}}\Psi_n(\del_\eta)\text{Ai}(\eta),\\
 \del_x\psi&=\sum_{n=0}^\infty \hbar^{\frac{2n}{3}+\frac{1}{6}}\til{\Psi}_n(\del_\eta)\text{Ai}(\eta),
\end{aligned} 
\end{equation}
where $\eta=\hbar^{-\frac{2}{3}}z^2$. By matching the WKB expansion of $\psi,\del_x\psi$
and the asymptotic expansion of the Airy function, we find the first few terms of 
$\Psi_n,\til{\Psi}_n$
\begin{equation}
\begin{aligned}
 \Psi_0&=1,\quad &\til{\Psi}_0=&-\del_\eta,\\
\Psi_1&=\del_\eta^2-\frac{4}{15}\del_\eta^5,\quad &\til{\Psi}_1=&
\frac{1}{2}-\frac{5\del_\eta^3}{3}+\frac{4 \del_\eta^6}{15},\\
\Psi_2&=-\frac{9 \del_\eta}{8}+\frac{5\del_\eta^4}{2}
-\frac{212\del_\eta^7}{315}+\frac{8 \del_\eta^{10}}{225},\quad &\til{\Psi}_2=&
\frac{33 \del_\eta^2}{8}-\frac{9\del_\eta^5}{2}+\frac{268\del_\eta^8}{315}-\frac{8\del_\eta^{11}}{225}.
\end{aligned} 
\end{equation}
Above $\Psi_n$ agrees with the result in our previous paper \cite{Okuyama:2019xbv} 
obtained by a different method.

\subsection{Trace formula}

In \cite{Okuyama:2019xbv} we showed that
the one-boundary partition function is expressed as
\begin{align}
\label{eq:Z1Tr}
\begin{aligned}
Z_1(\beta)
 &=\int_{-\infty}^x dx'\langle x'|e^{\beta Q}|x'\rangle
  =\Tr\left[e^{\beta Q}\Pi\right],
\end{aligned}
\end{align}
where we have introduced the projector
\begin{align}
\Pi =\int_{-\infty}^x dx'|x'\rangle\langle x'|.
\end{align}
As shown in \cite{Okuyama:2018aij} the general connected correlator
is given by
\begin{align}
\begin{aligned}
Z_n(\{\beta_i\},\{t_k\})
 &=\Tr\log
   \left(1+\left[-1+\prod_{i=1}^n(1+z_i e^{\beta_iQ})\right]\Pi\right)
   \Bigg|_{{\cal O}(z_1\cdots z_n)}\\
 &=\Tr\log
   \left(1+\sum_{k=1}^n\sum_{i_1<\cdots<i_k}z_{i_1}\cdots z_{i_k}
           e^{(\beta_{i_1}+\cdots\beta_{i_k})Q}\Pi\right)
   \Bigg|_{{\cal O}(z_1\cdots z_n)}.
\end{aligned}
\end{align}
For instance, two- and three-boundary correlators are 
written explicitly as
\begin{align}
Z_2(\beta_1,\beta_2)
 &=\Tr\left[e^{(\beta_1+\beta_2)Q}\Pi
   -e^{\beta_1 Q}\Pi e^{\beta_2 Q}\Pi\right],
\label{eq:corr-Pi}
\\
Z_3(\beta_1,\beta_2,\beta_3)
 &=\Tr\left[e^{(\beta_1+\beta_2+\beta_3)Q}\Pi
   +e^{\beta_1 Q}\Pi e^{\beta_2 Q}\Pi e^{\beta_3 Q}\Pi
   +e^{\beta_1 Q}\Pi e^{\beta_3 Q}\Pi e^{\beta_2 Q}\Pi\right.\nonumber\\
 &\hspace{2.7em}\left.
   -e^{\beta_1 Q}\Pi e^{(\beta_2+\beta_3) Q}\Pi
   -e^{\beta_2 Q}\Pi e^{(\beta_3+\beta_1) Q}\Pi
   -e^{\beta_3 Q}\Pi e^{(\beta_1+\beta_2) Q}\Pi\right].
\end{align}
In general, $Z_n$ is a sum of the multi-boundary correlator
\begin{align}
\label{eq:cKdef}
\Tr(e^{\beta_1 Q}\Pi \cdots e^{\beta_n Q}\Pi)
 =:\exp\left[{\cK^{(n)}(\beta_1,\ldots,\beta_n)}\right].
\end{align}
In \cite{Okuyama:2019xbv} we saw that
$\cK^{(1)}=\cF$ admits the 't Hooft expansion
$\cF=\sum_{k=0}^\infty\hbar^{k-1}\cF_k$.
Similarly, in what follows we explicitly show
that $\cK^{(n)}$ admits the 't Hooft expansion
\begin{align}
\cK^{(n)}=\sum_{k=0}^\infty\hbar^{k-1}\cK^{(n)}_k.
\end{align}
%

\subsection{Darboux-Christoffel kernel}

Let $|\xi\rangle$ be the energy eigenstate 
corresponding to $\psi(\xi)$ in section~\ref{sec:BAgexp},
namely
\begin{align}
\begin{aligned}
Q|\xi\rangle &= \xi|\xi\rangle,\\
\langle x|\xi\rangle&=\langle\xi|x\rangle=\psi(\xi,t_0=\hbar x).
\end{aligned}
\end{align}
$|\xi\rangle$ is normalized so that
\begin{align}
\label{eq:Eid}
1=\int_{-\infty}^\infty d\xi |\xi\rangle \langle \xi|.
\end{align}
By inserting $n$ copies of \eqref{eq:Eid} with
variables $\xi_i\ (i=1,\ldots,n)$,
the multi-boundary correlator is expressed as
\begin{align}
\label{eq:nptint}
\begin{aligned}
e^{\cK^{(n)}}
&=\Tr(e^{\beta_1 Q}\Pi \cdots e^{\beta_n Q}\Pi)\\
&=
  \int_{-\infty}^\infty d\xi_1\cdots\int_{-\infty}^\infty d\xi_n\,
 e^{\sum_{i=1}^n\beta_i\xi_i}K_{12}K_{23}\cdots K_{n1},
\end{aligned}
\end{align}
where
\begin{align}
\begin{aligned}
K_{ij}\equiv K(\xi_i,\xi_j)
 &=\langle \xi_i|\Pi|\xi_j\rangle\\
 &=\int_{-\infty}^x dx'\langle \xi_i|x'\rangle\langle x'|\xi_j\rangle
  =\int_{-\infty}^x dx'\psi(\xi_i)\psi(\xi_j)
\end{aligned}
\end{align}
is the Darboux-Christoffel kernel. Since $\psi(\xi)$
satisfies the Schr\"odinger equation \eqref{eq:Schrxi},
we see that
\begin{align}
\begin{aligned}
(\xi_i-\xi_j)K_{ij}
 &=\int_{-\infty}^x dx'
  \left[\left[\left(\partial_{x'}^2+u\right)\psi(\xi_i)\right]\psi(\xi_j)
  -\psi(\xi_i)\left(\partial_{x'}^2+u\right)\psi(\xi_j)\right]\\
 &=\int_{-\infty}^x dx'
  \partial_{x'}
  \left[\partial_{x'}\psi(\xi_i)\psi(\xi_j)
  -\psi(\xi_i)\partial_{x'}\psi(\xi_j)\right]\\
 &=\partial_x\psi(\xi_i)\psi(\xi_j)
   -\psi(\xi_i)\partial_x\psi(\xi_j).
\end{aligned}
\end{align}
The Darboux-Christoffel kernel then becomes
\begin{align}
\begin{aligned}
K_{ij}
 &=\frac{\partial_x\psi(\xi_i)\psi(\xi_j)
        -\psi(\xi_i)\partial_x\psi(\xi_j)}
        {\xi_i-\xi_j}\\
 &=e^{A(\xi_i)+A(\xi_j)}
  \frac{\partial_x A(\xi_i)-\partial_x A(\xi_j)}{\xi_i-\xi_j}
  =e^{A(\xi_i)+A(\xi_j)}\frac{v(\xi_i)-v(\xi_j)}{\xi_i-\xi_j}.
\end{aligned}
\end{align}
Plugging this expression into \eqref{eq:nptint}
and using the genus expansion of $A(\xi)$
calculated in section~\ref{sec:BAgexp},
one can compute the 't Hooft expansion of $\cK^{(n)}$,
as we see below.

\subsection{Saddle point calculation}\label{sec:saddle}

In \cite{Okuyama:2019xbv} we considered the 't Hooft expansion
of $\cF=\log Z_\JT$
\begin{align}
\cK^{(1)}=\cF=\sum_{k=0}^\infty\hbar^{k-1}\cF_k
\end{align}
and calculated the first three coefficients
$\cF_k\ (k=0,1,2)$ at the on-shell value $(y,t)=(0,1)$.
In what follows let us generalize the calculation
to the off-shell as well as the multi-boundary cases.

Let us first consider the off-shell generalization of the
above free energy.
Using the technique developed in the previous sections we have
\begin{align}
\begin{aligned}
e^{\cF}
 &=
   \int_{-\infty}^\infty d\xi e^{\frac{\lambda\xi}{\hbar}}
  K(\xi,\xi)\\
 &=
   \int_{-\infty}^\infty d\xi e^{\frac{\lambda\xi}{\hbar}+2A(\xi)}
   \partial_\xi v(\xi)
  =
   \int_{-\infty}^\infty d\xi e^{\frac{\lambda\xi}{\hbar}+2A(\xi)}
   \frac{1}{2z}\partial_z v(\xi)\\
 &=
   \int_{-\infty}^\infty d\xi
  e^{\left[\lambda\xi+2A_0(\xi)\right]\hbar^{-1}
    +2A_1(\xi)+ 2A_2(\xi)\hbar+{\cal O}(\hbar^2)}
  \frac{1}{2z}\left(1+\partial_z v_1(\xi)\hbar+{\cal O}(\hbar^2)\right).
\end{aligned}
\label{eq:cFpathint}
\end{align}
The above integral is evaluated by the saddle point approximation.
The saddle point $\xi_*$ is given by the condition
\begin{align}
\label{eq:saddlecond}
\partial_\xi\left[\lambda\xi+2A_0(\xi)\right]\Big|_{\xi=\xi_*}=0.
\end{align}
This is equivalent to
\begin{align}
\label{eq:laexp}
\begin{aligned}
\lambda
 &=\partial_\xi V_\eff(\xi)\Big|_{\xi=\xi_*}\\
 &=2tz_*+\sum_{n=1}^\infty\frac{(-1)^n n!B_n}{(2n+1)!}(2z_*)^{2n+1},
\end{aligned}
\end{align}
where $z_\ast := \sqrt{\xi_\ast-y}$
and $V_\eff(\xi)$ is the off-shell effective potential
defined in \eqref{eq:off-Veff}.
Inverting this relation we obtain
\begin{align}
\label{eq:zxiexp}
\begin{aligned}
z_*(\lambda)
 &=\frac{1}{2t}\lambda+\frac{B_1}{12t^4}\lambda^3
  +\left(\frac{B_1^2}{24t^7}-\frac{B_2}{120t^6}\right)\lambda^5
  +{\cal O}(\lambda^7),\\
\xi_*(\lambda)
 &=y+\frac{1}{4t^2}\lambda^2
 +\frac{B_1}{12t^5}\lambda^4
 +\left(\frac{7B_1^2}{144t^8}-\frac{B_2}{120t^7}\right)\lambda^6
 +{\cal O}(\lambda^8).
\end{aligned}
\end{align}
As in \cite{Okuyama:2019xbv} let us introduce a new variable $\phi$ as
\begin{align}
\xi=\xi_*+\sqrt{\hbar}\phi.
\end{align}
The integral \eqref{eq:cFpathint} is then written as
\begin{align}
\begin{aligned}
e^{\cF}
 &=
  e^{\left[\lambda\xi_*+2A_0(\xi_*)\right]\hbar^{-1}+2A_1(\xi_*)}
   \frac{1}{2z_*}
   \int_{-\infty}^\infty \sqrt{\hbar}d\phi
  e^{\partial_{\xi_*}^2A_0(\xi_*)\phi^2}
  \left(1+{\cal O}(\hbar)\right).
\end{aligned}
\end{align}
By expanding the integrand in $\hbar$,
the integral in $\phi$ can be performed as Gaussian integrals.
One can in principle calculate $\cF_k$ up to any order.
Evaluating the integral up to ${\cal O}(\hbar)$
for instance, we obtain
\begin{align}
\label{eq:offFn}
\begin{aligned}
\cF_0
 &=\lambda\xi_*+2A_0(\xi_*)
  =\lambda\xi_*-\int_y^{\xi_*}d\xi_*'\lambda(\xi_*')\\
 &=\int_0^\lambda d\lambda'\xi_\ast(\lambda'),\\
\cF_1
 &=
   2A_1(\xi_*)
   -\log(2z_*)
   +\frac{1}{2}\log\frac{\pi\hbar}{-\partial_{\xi_*}^2A_0(\xi_*)}\\
 &=\frac{1}{2}\log(\partial_\lambda\xi_*)-2\log z_*
  +\frac{1}{2}\log\frac{\hbar}{32\pi},\\[1ex]
\cF_2
 &=\frac{5(\partial_{\xi_*}^2\lambda)^2}{24(\partial_{\xi_*}\lambda)^3}
   -\frac{\partial_{\xi_*}^3\lambda}{8(\partial_{\xi_*}\lambda)^2}
   +\frac{\partial_{\xi_*}^2\lambda}{2z_*^2(\partial_{\xi_*}\lambda)^2}
   +\frac{1}{z_*^4\partial_{\xi_*}\lambda}
   -\frac{17}{24tz_*^3}-\frac{B_1}{12t^2z_*}.
\end{aligned}
\end{align}
In particular,
$\cF_0(\la)$ is given by the Legendre transform of
the effective potential $V_\eff(\xi)=-2A_0(\xi)$.
By using \eqref{eq:laexp} and \eqref{eq:zxiexp}
we see that they are expanded as
\begin{align}
\label{eq:offFnexp}
\begin{aligned}
\cF_0(\lambda)
 &=y\lambda+\frac{\lambda^3}{12t^2}+\frac{B_1\lambda^5}{60t^5}
   +\left(\frac{B_1^2}{144t^8}-\frac{B_2}{840t^7}\right)\lambda^7\\
&\hspace{1em}
   +\left(\frac{5B_1^3}{1296t^{11}}-\frac{B_1B_2}{720t^{10}}
     +\frac{B_3}{15120t^9}\right)\lambda^9
   +{\cal O}(\lambda^{11}),\\
\cF_1(\lambda)
 &=\frac{1}{2}\log\frac{\hbar}{4\pi}+\log t-\frac{3}{2}\log\lambda\\
&\hspace{1em}
 +\left(\frac{B_1^2}{24t^6}-\frac{B_2}{60t^5}\right)\lambda^4
 +\left(\frac{5B_1^3}{108t^9}
   -\frac{B_1B_2}{36t^8}+\frac{B_3}{420t^7}\right)\lambda^6
 +{\cal O}(\lambda^8),\\
\cF_2(\lambda)
 &=\frac{B_1}{t\lambda}-\frac{B_2}{12t^3}\lambda
   +\left(\frac{B_1^3}{18t^7}-\frac{29B_1B_2}{360t^6}
     +\frac{B_3}{72t^5}\right)\lambda^3+{\cal O}(\lambda^5).
\end{aligned}
\end{align}

At the on-shell value $(y,t)=(0,1)$
\eqref{eq:laexp} and \eqref{eq:zxiexp} reduce to
\begin{align}
\lambda=\sin(2z_*)\quad
\Leftrightarrow\quad
z_*=\sqrt{\xi_*}=\frac{1}{2}\arcsin\la
\end{align}
and the results \eqref{eq:offFn} reproduce
those obtained in \cite{Okuyama:2019xbv}:
\begin{align}
\label{eq:onFn}
\begin{aligned}
\cF_0(\lambda)
 &=\frac{1}{4}\lambda\arcsin(\lambda)^2
   +\frac{1}{2}\left(\sqrt{1-\lambda^2}\arcsin\lambda-\lambda\right),\\
\cF_1(\lambda)
 &=-\frac{3}{2}\log\arcsin\lambda
   -\frac{1}{4}\log(1-\lambda^2)
   +\frac{1}{2}\log\frac{\hbar}{4\pi},\\
\cF_2(\lambda)
 &=\frac{17}{3\arcsin(\lambda)^3}
   \left[-1+\frac{1}{\sqrt{1-\lambda^2}}\right]
   -\frac{23\lambda}{12(1-\lambda^2)\arcsin(\lambda)^2}\\
 &+\frac{1}{12\arcsin(\lambda)}
   \left[-2-\frac{2}{\sqrt{1-\lambda^2}}
         +\frac{5}{(1-\lambda^2)^{3/2}}\right].
\end{aligned}
\end{align}

In the same manner as above one can calculate
the 't Hooft expansion of the two-boundary correlator.
We start from
\begin{align}
\begin{aligned}
e^{\cK^{(2)}}
 &=
   \int_{-\infty}^\infty d\xi_1\int_{-\infty}^\infty d\xi_2\,
   e^{\frac{\lambda_1\xi_1+\lambda_2\xi_2}{\hbar}}
   e^{2A(\xi_1)+2A(\xi_2)}
   \left(\frac{v(\xi_1)-v(\xi_2)}{\xi_1-\xi_2}\right)^2.
\end{aligned}
\end{align}
It is clear that the saddle point $\xi_{i*}\ (i=1,2)$ is given by
\begin{align}
\label{eq:saddlecond2}
\partial_{\xi_i}\left[\lambda_i\xi_i+2A_0(\xi_i)\right]
\Big|_{\xi_i=\xi_{i*}}=0.
\end{align}
This is the same relation as in
the one-boundary case \eqref{eq:saddlecond}
and thus $\lambda_i$ and $\xi_i$ are related as in
\eqref{eq:laexp}--\eqref{eq:zxiexp}.
Evaluating the integral by the saddle point approximation
we find
\begin{align}
\label{eq:offKn}
\begin{aligned}
\cK_0^{(2)}
 &=\sum_{i=1}^2\left[\lambda\xi_{i*}+2A_0(\xi_{i*})\right]
  =\sum_{i=1}^2\cF_0(\lambda_i),\\
\cK_1^{(2)}
 &=\sum_{i=1}^2\left[
    2A_1(\xi_{i*})
    +\frac{1}{2}
     \log\frac{\pi\hbar}{-\partial_{\xi_{i*}}^2 A_0(\xi_{i*})}
               \right]
  -2\log(z_{1*}+z_{2*})\\
 &=\sum_{i=1}^2\left[
    \frac{1}{2}\log(\partial_{\lambda_i}\xi_{i*})-\log z_{i*}
    +\frac{1}{2}\log\frac{\hbar}{8\pi}
               \right]
  -2\log(z_{1*}+z_{2*}),\\
\cK_2^{(2)}
 &=\Biggl[
   \frac{5(\partial_{\xi_{1*}}^2\lambda_1)^2}
        {24(\partial_{\xi_{1*}}\lambda_1)^3}
   -\frac{\partial_{\xi_{1*}}^3\lambda_1}
         {8(\partial_{\xi_{1*}}\lambda_1)^2}
   +\frac{(3z_{1*}+z_{2*})\partial_{\xi_{1*}}^2\lambda_1}
         {4z_{1*}^2(z_{1*}+z_{2*})(\partial_{\xi_{1*}}\lambda_1)^2}\\
 &\hspace{2em}
   +\frac{3(5z_{1*}^2+4z_{1*}z_{2*}+z_{2*}^2)}
         {8z_{1*}^4(z_{1*}+z_{2*})^2\partial_{\xi_{1*}}\lambda_1}
   -\frac{12z_{1*}+5z_{2*}}{24tz_{1*}^3z_{2*}}
   -\frac{B_1}{12t^2z_{1*}}
   \Biggr]+(\lambda_1\leftrightarrow \lambda_2),
\end{aligned}
\end{align}
where $\cF_0$ is given in \eqref{eq:offFn}.

At the on-shell value $(y,t)=(0,1)$
the above results reduce to
\begin{align}
\label{eq:onKn}
\begin{aligned}
\cK_0^{(2)}
 =&\,\sum_{i=1}^2\left[\frac{1}{4}\lambda_i\arcsin(\lambda_i)^2
   +\frac{1}{2}
    \left(\sqrt{1-\lambda_i^2}\arcsin\lambda_i-\lambda_i\right)
   \right],\\
\cK_1^{(2)}
 =&\,
 -\frac{1}{4}\log[(1-\lambda_1^2)(1-\lambda_2^2)]
 -\frac{1}{2}\log(\arcsin\lambda_1\arcsin\lambda_2)\\
 &\,-2\log(\arcsin\lambda_1+\arcsin\lambda_2)
 +\log\frac{\hbar}{\pi},\\
\cK_2^{(2)}
 =&\,\sum_{k=0}^2\frac{f_k(\lambda_1)+f_k(\lambda_2)}
 {(\arcsin\lambda_1+\arcsin\lambda_2)^k}
\quad\mbox{with}\\
&f_0(\lambda)
 =\frac{5}{3\arcsin(\lambda)^3}
   \left[-1+\frac{1}{\sqrt{1-\lambda^2}}\right]
   -\frac{11\lambda}{12(1-\lambda^2)\arcsin(\lambda)^2}\\
 &\hspace{4em}
  +\frac{1}{12\arcsin(\lambda)}
   \left[-2-\frac{2}{\sqrt{1-\lambda^2}}
         +\frac{5}{(1-\lambda^2)^{3/2}}\right],\\
&f_1(\lambda)
 =
 -\frac{2\lambda}{(1-\lambda^2)\arcsin\lambda}
 -\left(1-\frac{1}{\sqrt{1-\lambda^2}}\right)
  \frac{4}{\arcsin(\lambda)^2},\\
&f_2(\lambda)
 =\left(-8+\frac{6}{\sqrt{1-\lambda^2}}\right)\frac{1}{\arcsin\lambda}.
\end{aligned}
\end{align}

In the same way one can calculate $\cK_k^{(n)}$
for $n\ge 3$.
We find that $\cK_{0,1}^{(n)}(\lambda_1,\ldots,\lambda_n)$
for $n\in\bbZ_{>0}$ take the universal form
\begin{align}
\begin{aligned}
\cK_0^{(n)}(\lambda_1,\ldots,\lambda_n)
 &=\sum_{i=1}^n\cF_0(\lambda_i),\\
\cK_1^{(n)}(\lambda_1,\ldots,\lambda_n)
 &=\sum_{i=1}^n\left[
    \frac{1}{2}\log(\partial_{\lambda_i}\xi_{i*})-\log z_{i*}
    +\frac{1}{2}\log\frac{\hbar}{8\pi}
               \right]
  -\sum_{i=1}^n\log(z_{i*}+z_{i+1,*}),
\end{aligned}
\end{align}
where the subscript $i$ should be identified mod $n$.

We also find that $\cK_2^{(3)}$ at the on-shell value is given by
\begin{align}
\begin{aligned}
&\cK^{(3)}_2(\lambda_1,\lambda_2,\lambda_3)\Big|_{y=0,t=1}\\
&=\sum_{i=1}^3\left[\cF_2(\lambda_i)+\frac{1}{2(z_{i*})^3}\right]\\
&\hspace{.5em}
 +\left[\left(
   \frac{3(z_{1*})^4+2(z_{2*}+z_{3*})(z_{1*})^3-3z_{2*}z_{3*}(z_{1*})^2
         -3z_{2*}z_{3*}(z_{2*}+z_{3*})z_{1*}-2(z_{2*})^2(z_{3*})^2}
        {4(z_{1*})^3(z_{1*}+z_{2*})^2(z_{1*}+z_{3*})^2\cos(2z_{1*})}
  \right.\right.\\
&\hspace{1.5em}
    \left.\left.\phantom{\frac{1}{1}}
   -\frac{z_{1*}+z_{2*}}{4(z_{1*}z_{2*})^2}
   +\frac{z_{2*}z_{3*}-(z_{1*})^2}
         {4(z_{1*})^2(z_{1*}+z_{2*})(z_{1*}+z_{3*})}
   \frac{\sin(2z_{1*})}{\cos^2(2z_{1*})}\right)
    +\mbox{cyclic perm.}\right],
\end{aligned}
\end{align}
where $\cF_2$ at the on-shell value is given in \eqref{eq:onFn}.

\section{Low temperature expansion of two-boundary correlator}\label{sec:lowT}

In this section let us consider the low temperature expansion
of the two-boundary correlator.
More specifically, we consider the situation where
\begin{align}
\label{eq:Tdef}
T=\frac{1}{\beta}=\frac{1}{\beta_1+\beta_2}
\end{align}
is small and calculate the expansion of $Z_2(\beta_1,\beta_2)$ in $T$.
To begin with, one can observe that
the leading order term in each coefficient
$Z_{g,2}$ \eqref{eq:Zg2results}
of the genus expansion \eqref{eq:Z2gexp}
is independent of $y$
and has the form ${\cal O}(\beta^{3g-1}t^{-2g})$.
We find that they can be summed over genus
\begin{align}
\label{eq:TexpZ2leading}
\begin{aligned}
&\hspace{-1em}
\frac{\sqrt{\beta_1\beta_2}}{2\pi}e^{(\beta_1+\beta_2)y}
 \left[\frac{1}{\beta_1+\beta_2}
  +\frac{\beta_1^2+\beta_1\beta_2+\beta_2^2}{24t^2}\gs^2+\cdots\right]\\
 &=\frac{t}{2\sqrt{\pi}\hbar(\beta_1+\beta_2)^{3/2}}
   e^{(\beta_1+\beta_2)y+\frac{\hbar^2(\beta_1+\beta_2)^3}{12t^2}}
   \Erf\left(\frac{\hbar}{2t}\sqrt{\beta_1\beta_2(\beta_1+\beta_2)}
       \right)\\
 &=\frac{t}{2\sqrt{\pi}h}
   e^{\frac{y}{T}+\frac{h^2}{12t^2}}
   \Erf\left(\frac{h}{2t}\sqrt{r(1-r)}\right).
\end{aligned}
\end{align}
Here $\Erf(z)$ is the error function
\begin{align}
\Erf(z)=\frac{2}{\sqrt{\pi}}\int_0^z dt e^{-t^2}
\end{align}
and we have introduced the notation
\begin{align}
h=\hbar \beta^{3/2},\quad
r=\frac{\beta_1}{\beta}.
\end{align}
At the on-shell value $(y,t)=(0,1)$
\eqref{eq:TexpZ2leading} precisely reproduces
the result of the Airy case \cite{okounkov2002generating}
(discussed also in our previous paper \cite{Okuyama:2019xbv}).
More generally, including the subleading corrections
the two-point function at the on-shell value
\eqref{eq:Z2decomp}--\eqref{eq:gexpZ2formal} is written as
\begin{equation}
\begin{aligned}
 \bra Z(\bt_1)Z(\bt_2)\ket_\conn=\frac{\rt{r(1-r)}}{2\pi}\left[1+
\sum_{g=1}^\infty (\rt{2}h)^{2g}\int_{\b{\mathcal{M}}_{g,2}}\frac{e^{T\ka}}
{(1-r\psi_1)(1-(1-r)\psi_2)}\right].
\end{aligned} 
\label{eq:two-kappa}
\end{equation}

Regarding the above result,
as in the case of $Z_1$ \cite{Okuyama:2019xbv}
it is natural to make an ansatz
\begin{align}
\label{eq:TexpZ2anz1}
Z_2(\beta_1,\beta_2)
 =\frac{e^{\frac{y}{T}+\frac{h^2}{12t^2}}}{2\sqrt{\pi}h}
  \sum_{\ell=0}^\infty\frac{T^\ell}{\ell!}z^{(2)}_\ell
\end{align}
with
\begin{align}
z^{(2)}_0=t\Erf\left(\frac{h}{2t}\sqrt{r(1-r)}\right).
\end{align}
The subleading parts $z^{(2)}_\ell\ (\ell\ge 1)$ can
also be estimated from the data of
the genus expansion \eqref{eq:Z2gexp}, \eqref{eq:Zg2results}.
We find that $z^{(2)}_\ell$ has the structure
\begin{align}
\label{eq:TexpZ2anz2}
z^{(2)}_\ell(\beta_1,\beta_2)
 =\Erf\left(\frac{h}{2t}\sqrt{r(1-r)}\right)z_\ell(\beta)
  +h\sqrt{\frac{r(1-r)}{\pi}}e^{-\frac{h^2r(1-r)}{4t^2}}
   g_\ell(\beta_1,\beta_2)
\end{align}
with
\begin{align}
\label{eq:zlgl}
\begin{aligned}
z_0&=t,\quad z_1=\left(1+\frac{h^4}{60t^4}\right)B_1,\quad\ldots,\\
g_0&=0,\quad
g_1=\left(-\frac{1}{t}+\frac{h^2(1-r+r^2)}{6t^3}\right)B_1,\quad,\ldots.
\end{aligned}
\end{align}
Interestingly, the above $z_\ell$ coincides with the coefficient
$z_\ell$ of the low temperature expansion of $Z_1$
studied in \cite{Okuyama:2019xbv}.
In \cite{Okuyama:2019xbv} we saw that
$z_\ell$ can be calculated by solving a set of recursion relations
following from the KdV constraint.
Similarly, one can derive recursion relations for $g_\ell$,
as we will see below. We should emphasize that from \eqref{eq:two-kappa}
$z_\ell$ and $g_\ell$ contain the all-genus
information of the intersection numbers
at the fixed power of $\ka$, i.e. $\ka^\ell$.

Let us first express the small $T$ expansion of $Z_2$ as
\begin{align}
\label{eq:TexpZ2}
Z_2(\beta_1,\beta_2)
 &=\cA\sum_{\ell=0}^\infty\frac{T^{\ell+1}}{\ell!}z_\ell
  +\cB\sum_{\ell=0}^\infty\frac{T^{\ell+1}}{\ell!}g_\ell,
\end{align}
where
\begin{align}
\cA
 =\frac{1}{2\sqrt{\pi} h T}
   e^{\frac{y}{T}+\frac{h^2}{12t^2}}
   \Erf\left(\frac{h}{2t}\sqrt{r(1-r)}\right),\quad
\cB
 =\frac{\sqrt{r(1-r)}}{2\pi T}
   e^{\frac{y}{T}+\frac{h^2r^3}{12t^2}+\frac{h^2(1-r)^3}{12t^2}}.
\end{align}
Using the properties
\begin{align}
\label{eq:ABdiffrel}
\partial_t \cA
 =-\frac{h^2}{6t^3}\cA-\frac{1}{t^2}\cB,\qquad
\partial_t \cB
 =-\frac{h^2[r^3+(1-r)^3]}{6t^3}\cB
\end{align}
it is not difficult to see that
the small $T$ expansion of $W_2=\partial_x Z_2$ takes the form
\begin{align}
\label{eq:TexpW2}
W_2(\beta_1,\beta_2)
 &=\cA\sum_{\ell=0}^\infty T^\ell w_\ell(h)
  +\cB\sum_{\ell=0}^\infty T^\ell b_\ell,
\end{align}
where $w_\ell(h)$ is the expansion coefficient for $W_1(\beta)$
introduced in \cite{Okuyama:2019xbv} and
$b_\ell$ is some polynomial in $h,r,t^{-1},B_n\ (n\ge 1)$.
The small $T$ expansions of $W_1(\beta_1)$ and $W_1(\beta_2)$ are
explicitly written as
\begin{align}
\label{eq:TexpW1W1}
\begin{aligned}
W_1(\beta_1)
 &=e^{\frac{ry}{T}+\frac{h^2r^3}{12t^2}}
   \sum_{\ell=0}^\infty\left(\frac{T}{r}\right)^{\ell+1}
   w_\ell(h r^{3/2}),\\
W_1(\beta_2)
 &=e^{\frac{(1-r)y}{T}+\frac{h^2(1-r)^3}{12t^2}}
   \sum_{\ell=0}^\infty\left(\frac{T}{1-r}\right)^{\ell+1}
   w_\ell(h(1-r)^{3/2}).
\end{aligned}
\end{align}
As we saw in \cite{Okuyama:2019xbv},
$w_\ell(h)$ is
determined by the KdV constraint
for the one-point function
\begin{align}
\label{eq:W1KdVconst}
\begin{aligned}
-\partial_t W_1
 &=\hat{u}\partial_0 W_1
  +\frac{h^2T^3}{6}\partial_0^3 W_1,\\
\hat{u}
 &=u-y=\sum_{g=1}^\infty 2^g h^{2g} T^{3g} u_g.
\end{aligned}
\end{align}
Similarly, $b_\ell$ can be computed from the KdV constraint
for the two-point function
\begin{align}
\label{eq:W2KdVconst}
-\partial_t W_2
 =\hat{u}\partial_0 W_2
 +\frac{h^2T^3}{6}\partial_0^3 W_2+\partial_0 W_1\partial_0 W_1
\end{align}
by equating the terms proportional to $\cB$.
Note that the last term of \eqref{eq:W2KdVconst}
is proportional to $\cB$.
If we formally set $\cB=0$, we obtain the homogeneous equation for $W_2$
which is equivalent to the KdV constraint for the one-point
function \eqref{eq:W1KdVconst}.
This justifies the ansatz \eqref{eq:TexpW2}
for $W_2$
(and thus our original conjectures
\eqref{eq:TexpZ2anz1} and \eqref{eq:TexpZ2anz2} for $Z_2$)
where the expansion coefficient of the first term
is given by that of the one-point function $w_\ell$.

Plugging \eqref{eq:TexpW2}--\eqref{eq:TexpW1W1}
into \eqref{eq:W2KdVconst}
and using the relations \eqref{eq:ABdiffrel}
one finds that the recursion relation for $b_\ell$ is given by
\begin{align}
\begin{aligned}
&-\partial_t b_\ell
 +\frac{h^2}{6t^3}\left[r^3+(1-r)^3-1\right]b_\ell
 +\frac{1}{t^2}w_\ell(h)\\
&=\left(\hat{u}\partial_0+\frac{h^2T^3}{6}\partial_0^3\right)
 \left(\cA\sum_{j=0}^\ell T^jw_j(h)+\cB\sum_{j=1}^{\ell-1}T^jb_j\right)
 \Bigg|_{\cB,T^\ell}\\
&\hspace{1em}
 +\left(\sum_{j=0}^\ell\left(\frac{T}{r}\right)^{j+1}
  \left(\partial_0+\frac{r}{tT}+\frac{h^2r^3B_1}{6t^4}\right)
  w_j(hr^{3/2})\right)
 \cdot\left(r\to 1-r\right)\Bigg|_{T^\ell}.
\end{aligned}
\end{align}
Starting from $b_0=0$ one can compute $b_\ell$.
For instance, the first term is
\begin{align}
b_1=\frac{h^2B_1}{6t^4}(1-r+r^2).
\end{align}
From $\partial_x Z_2=W_2$, one can show that
\begin{align}
g_\ell
 =t\ell!b_\ell
 -t\ell\left[\partial_0 g_{\ell-1}
   +\frac{h^2B_1\left(r^3+(1-r)^3\right)}{6t^4}g_{\ell-1}
   +\frac{B_1}{t^3}z_{\ell-1}\right].
\end{align}
Starting from $g_0=0$ one can calculate $g_\ell\ (\ell\ge 1)$
up to arbitrary high order.
We have verified that this indeed reproduces
our conjectured results \eqref{eq:zlgl} estimated from
the genus expansion.

A few remarks are in order.
First, it is worth noting that
the two-point function admits
a low-temperature expansion of the form
\begin{align}
\label{eq:Z2ptinD}
Z_2(\beta_1,\beta_2)=\Erf(\sqrt{D})Z_1(\beta)
\end{align}
with
\begin{align}
D=\sum_{\ell=0}^\infty T^\ell D_\ell,\quad D_0=\frac{h^2}{4t^2}r(1-r).
\label{eq:Dell}
\end{align}
The structure of $Z_2(\bt_1,\bt_2)$ in \eqref{eq:TexpZ2}
is naturally understood from \eqref{eq:Z2ptinD}
by expanding the error function in $T$.
Note that \eqref{eq:Z1Tr} and \eqref{eq:Z2ptinD} imply that
the two terms in \eqref{eq:corr-Pi} correspond to
\begin{align}
\begin{aligned}
 \Tr (e^{(\bt_1+\bt_2)Q}\Pi)&=Z_1(\beta),\\
\Tr (e^{\bt_1Q}\Pi  e^{\bt_2Q}\Pi)&=\text{Erfc}(\sqrt{D})Z_1(\beta),
\end{aligned} 
\label{eq:tr-vs-Erfc}
\end{align}
where
\begin{align}
\Erfc(z) = 1-\Erf(z)
\end{align}
is the complementary error function.
To calculate $D_\ell$ in \eqref{eq:Dell},
it is convenient to introduce the normalized coefficients
$c_\ell:=D_\ell/D_0$ and expand $D$ as
\begin{align}
\label{eq:Dexpansion}
D=D_0\sum_{\ell=0}^\infty T^\ell c_\ell,
\quad c_0=1.
\end{align}
One can rewrite \eqref{eq:Z2ptinD} as
\begin{align}
\Erf(\sqrt{D_0})\sum_{\ell=0}^\infty\frac{T^\ell}{\ell!}z_\ell
 +2t\sqrt{\frac{D_0}{\pi}}e^{-D_0}
 \sum_{\ell=1}^\infty\frac{T^\ell}{\ell!}g_\ell
=\Erf(\sqrt{D})\sum_{\ell=0}^\infty\frac{T^\ell}{\ell!}z_\ell
\end{align}
or
\begin{align}
\begin{aligned}
\frac{\sum_{\ell=1}^\infty\frac{T^{\ell}}{\ell!}g_\ell}
     {\sum_{\ell=0}^\infty\frac{T^{\ell}}{\ell!}\hz_\ell}
&=\frac{1}{2}\sqrt{\frac{\pi}{D_0}}e^{D_0}
 \left(\Erf(\sqrt{D})-\Erf(\sqrt{D_0})\right)\\
&=\frac{1}{2}c_1T
 +\left(\frac{c_2}{2}-\frac{c_1^2}{8}-\frac{D_0c_1^2}{4}\right)T^2
 +{\cal O}(T^3),
\end{aligned}
\end{align}
where $\hz_\ell:=z_\ell/t$. (Note that $\hz_0=1$).
By comparing both sides of the equation
one can express $c_\ell$ in terms of
$\hz_\ell$ and $g_\ell$. First few of the results read
\begin{align}
\label{eq:cn}
\begin{aligned}
c_1&= 2g_1,\\
c_2&= 2D_0g_1^2+g_1^2-2g_1\hz_1+g_2,\\
c_3&=\frac{8}{3}D_0(D_0+1)g_1^3-2\hz_1(2D_0+1)g_1^2
 +(2D_0g_2+2\hz_1^2+g_2-\hz_2)g_1-\hz_1g_2+\frac{1}{3}g_3.
\end{aligned}
\end{align}
From these expressions one immediately obtains $D_\ell=D_0c_\ell$.

Second, as discussed in \cite{Okuyama:2019xbv},
given the result of the low temperature expansion
it is straightforward to take the 't Hooft limit \eqref{eq:tHooft}
and one can rearrange the low temperature expansion
as the 't Hooft expansion.
From the above results
one can compute the 't Hooft expansion of $D$
\begin{equation}
\begin{aligned}
 D=\sum_{n=0}^\infty\hbar^{n-1}\cD_n,
\end{aligned} 
\label{eq:Dexp}
\end{equation}
where $\cD_n$ is obtained as a double series expansion in
$(\lambda_1,\lambda_2)=(\hbar\bt_1,\hbar\bt_2)$.
Alternatively,
from the relation
\begin{equation}
\label{eq:DKrel}
\begin{aligned}
\Erfc(\rt{D})Z_1(\bt)
 =\Tr(e^{\beta_1 Q}\Pi e^{\beta_2 Q}\Pi)=e^{\mathcal{K}^{(2)}}
\end{aligned} 
\end{equation}
and the result of $\mathcal{K}^{(2)}$ in \eqref{eq:offKn}, 
one can calculate $\cD_n$
as exact functions.
Here, the complementary error function
can be expanded in $\hbar$ with the help of the asymptotic formula
\begin{align}
\label{eq:erfcexp}
\Erfc(z)
 =\frac{e^{-z^2}}{\sqrt{\pi}z}
  \sum_{n=0}^\infty\frac{(2n-1)!!}{(-2z^2)^n}.
\end{align}
For instance, the leading term is given by
\begin{equation}
\begin{aligned}
 \cD_0=\cF_0(\la_1+\la_2)-\cF_0(\la_1)-\cF_0(\la_2).
\end{aligned} 
\end{equation}
The higher order corrections $\cD_{n\geq1}$ can also be 
easily obtained from the result of $\mathcal{K}^{(2)}_{n\geq1}$
in \eqref{eq:offKn}.
We verified at the on-shell value $(y,t)=(0,1)$ that
the series expansions of $\cD_n$ obtained from \eqref{eq:cn}
are in perfect agreement with the exact expressions of
$\cD_n\ (n=1,2)$ obtained through \eqref{eq:DKrel}.

Third, one might think that 
$g_\ell$ would be interpreted as the expansion
coefficients for $\Tr(e^{\beta_1 Q}\Pi e^{\beta_2 Q}\Pi)$.
This intuition, however, is not precise. Rather, 
by using \eqref{eq:corr-Pi}, \eqref{eq:Z1Tr} and \eqref{eq:TexpZ2}
$\Tr(e^{\beta_1 Q}\Pi e^{\beta_2 Q}\Pi)$ is rewritten as
\begin{align}
\begin{aligned}
\Tr(e^{\beta_1 Q}\Pi e^{\beta_2 Q}\Pi)
&=Z_1(\beta)-Z_2(\beta_1,\beta_2)\\
&=\Erfc(\sqrt{D_0})Z_1(\beta)
  -\cB\sum_{\ell=0}^\infty\frac{T^{\ell+1}}{\ell!}g_\ell\\
&=\Erfc(\sqrt{D_0})
  \frac{e^{\frac{h^2}{12t^2}+\frac{y}{T}}}{2\sqrt{\pi}h}
  \sum_{\ell=0}^\infty\frac{T^\ell}{\ell!}z_\ell
  -\cB\sum_{\ell=0}^\infty\frac{T^{\ell+1}}{\ell!}g_\ell.
\end{aligned} 
\label{eq:TexpK2}
\end{align}
This clearly shows that not only $g_\ell$ but also $z_\ell$
are involved in the low temperature expansion of
$\Tr(e^{\beta_1 Q}\Pi e^{\beta_2 Q}\Pi)$.
By rearranging the low temperature expansion as
the 't Hooft expansion and using the asymptotic expansion formula
\eqref{eq:erfcexp},
we explicitly verified
at the on-shell value $(y,t)=(0,1)$ that
\eqref{eq:TexpK2} is indeed in agreement with
$e^{\cK^{(2)}}=e^{\sum_{k=0}^\infty \hbar^{k-1}\cK_k^{(2)}}$
with $\cK_k^{(2)}\ (k=0,1,2)$ given in \eqref{eq:onKn}.

\section{Spectral form factor in JT gravity}\label{sec:SFF}

In this section we will study the spectral form factor (SFF)
in JT gravity using our result of two-point function.
The SFF is extensively studied in the SYK model as a useful diagnostic of the late-time
chaos \cite{Garcia-Garcia:2016mno,Cotler:2016fpe,Saad:2018bqo,Saad:2019pqd}.
The SFF of chaotic system exhibits a
characteristic behavior called the ramp and the plateau.
From the bulk gravity perspective, the ramp comes from the Euclidean wormhole
connecting the two boundaries.
The plateau behavior, on the other hand, is a doubly non-perturbative
effect with respect to the Newton's constant whose bulk gravity interpretation is still
missing. From the random matrix model picture,
the origin of the plateau can be traced back to the
universal eigenvalue correlation given by the so-called sine-kernel formula.
However, this argument is based on the matrix model before taking the double-scaling limit
and the analytic form of the SFF in the JT gravity case has not been obtained 
yet as far as we know. Using our result in the previous section,
we can explicitly write down the analytic form of
SFF in JT gravity and see how the ramp and the plateau come about.

\begin{figure}[thb]
\centering
\includegraphics[width=8cm]{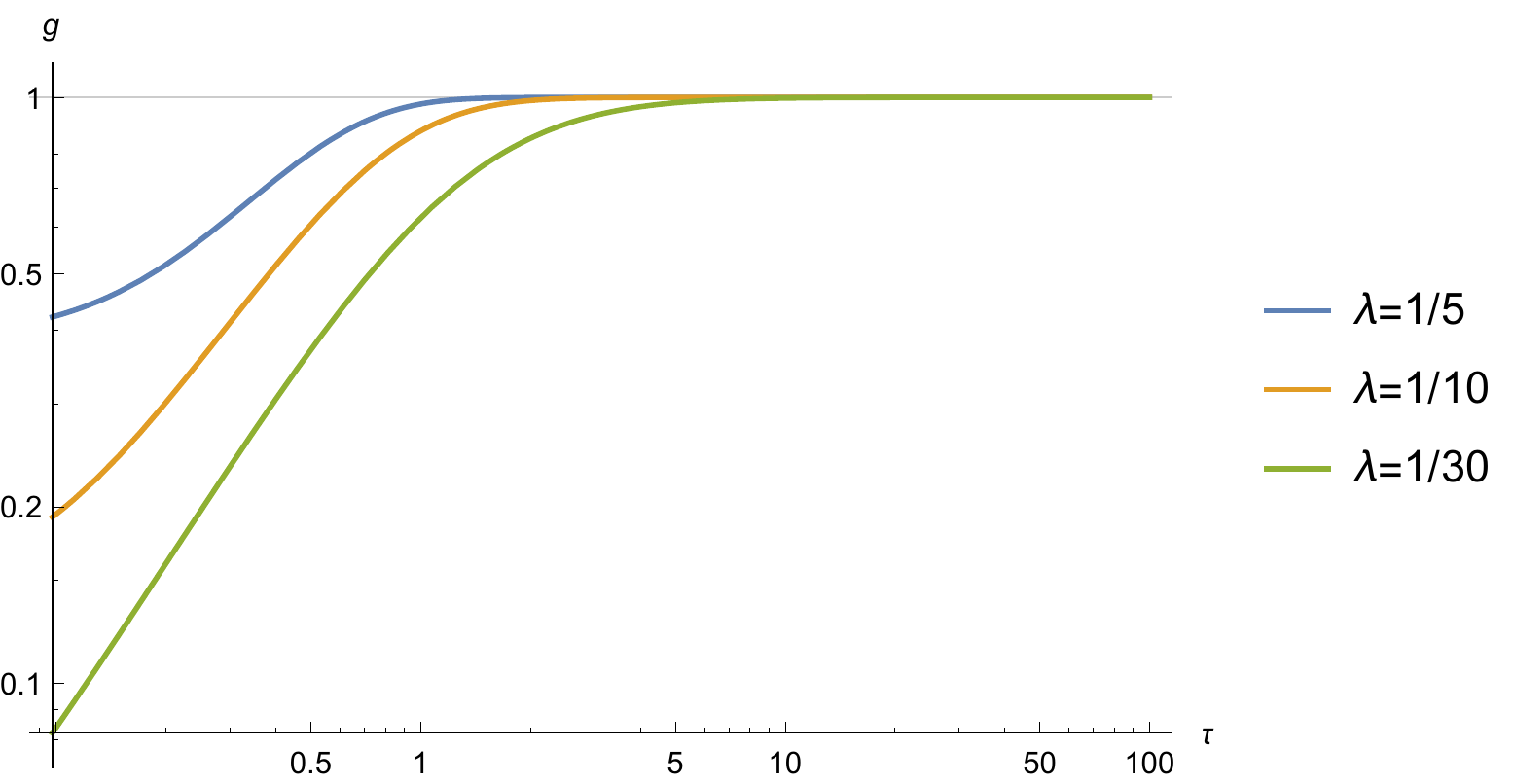}
  \caption{Plot of the spectral form factor $g(\bt,t,\hbar)$ as a function
of $\tau=\hbar t$ for $\hbar=1/30$.} 
  \label{fig:SFF}
\end{figure}

The SFF is defined by analytically continuing
the two-boundary correlator $\bra Z(\bt_1)Z(\bt_2)\ket_{\conn}$ to
a complex value of the boundary length 
$\bt_{1,2}=\bt\pm\ri t$. It is convenient to define the normalized
SFF  by
\begin{equation}
\begin{aligned}
 g(\bt,t,\hbar):=\frac{\bra Z(\bt+\ri t)Z(\bt-\ri t)\ket_{\conn}}{\bra Z(2\bt)\ket}.
\end{aligned} 
\end{equation}
Using our result in section \ref{sec:lowT}, this is given by the error function
\eqref{eq:Z2ptinD}
\begin{equation}
\begin{aligned}
 g(\bt,t,\hbar)=\text{Erf}\bigl(\rt{D}\bigr).
\end{aligned} 
\end{equation}
We are interested in the late-time behavior of the SFF at the timescale of order $t\sim\hbar^{-1}$.
To study this regime, it is natural to take
the 't Hooft limit\footnote{$t$ and $\tau$ in this section
should not be confused with those used in the previous sections.}
\begin{equation}
\begin{aligned}
 \hbar\to0,~ \bt\to\infty,~ t\to\infty,\quad\text{with}\quad \la=\hbar\bt,~\tau=\hbar t
~~\text{fixed}.
\end{aligned} 
\end{equation}
As we have seen in section \ref{sec:lowT}, $D$ is expanded as \eqref{eq:Dexp}
in this 't Hooft limit.
To see the behavior of the ramp and the plateau, it is sufficient to 
take the first term of the 't Hooft expansion
\begin{equation}
\begin{aligned}
 D\approx \hbar^{-1}\cD_0=\hbar^{-1}\bigl[\cF_0(2\la)-\cF_0(\la+\ri\tau)-\cF_0(\la-\ri\tau)\bigr],
\end{aligned} 
\label{eq:lead-D}
\end{equation}
where $\cF_0(\la)$ is given by \eqref{eq:onFn}.\footnote{If we replace $\cF_0(\la)$
by a cubic polynomial $\cF_0(\la)=\frac{\la^3}{12}$ we obtain the 
SFF for the Airy case
\begin{equation}
\begin{aligned}
 g_{\text{Airy}}=\text{Erf}\left(\hbar\rt{\frac{\bt(\bt^2+t^2)}{2}}\right).
\end{aligned} 
\label{eq:SFF-Airy}
\end{equation}
We stress that the SFF in JT gravity is not equal to the Airy case \eqref{eq:SFF-Airy}
and we start to see the deviation at the order $\cO(\la^5)$ 
as mentioned in the introduction.
}
In Fig.~\ref{fig:SFF}, we show the plot of SFF in the approximation \eqref{eq:lead-D}
for $\hbar=1/30$ with several different values of $\la$.
One can see that the SFF exhibits the characteristic feature
of the ramp and the plateau.
We observe that the timescale of the transition from ramp to plateau depends 
on $\la$ as in the pure topological gravity case 
 \cite{Okuyama:2019xbv}.

\begin{figure}[thb]
\centering
\includegraphics[width=8cm]{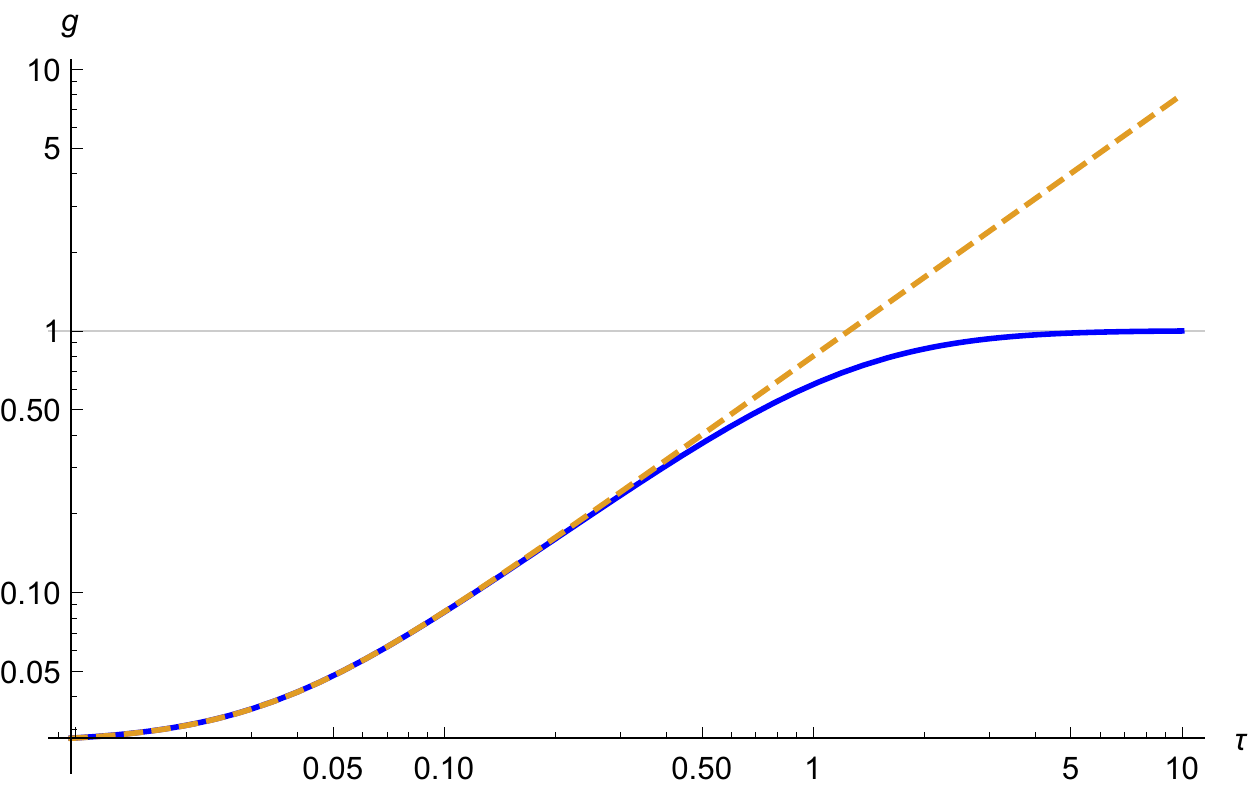}
  \caption{Plot of the spectral form factor 
for $\hbar=\la=1/30$. The orange dashed curve is the genus-zero result
while the blue solid curve is the plot of error function.} 
  \label{fig:SFF-genus0}
\end{figure}

In \cite{Saad:2019lba}
it is argued that the ramp is reproduced from the genus-zero part of the connected correlator
\begin{equation}
\begin{aligned}
 \bra Z(\bt+\ri t)Z(\bt-\ri t)\ket^{g=0}_{\conn}=\frac{\rt{\bt^2+t^2}}{4\pi\bt}=\frac{\rt{\la^2+\tau^2}}{4\pi\la}.
\end{aligned} 
\end{equation}
In Fig.~\ref{fig:SFF-genus0} we show the plot of the
genus-zero part (orange dashed curve) and the
full result (blue solid curve) for the SFF with $\hbar=\la=1/30$ as an example.
One can see that the genus-zero part captures the growing ramp behavior of SFF
at early times.
This agreement
at early times can be shown analytically
using the Taylor expansion of the error function and the
small $\la$ behavior of $\cF_0(\la)\sim\frac{\la^3}{12}$.

The appearance of the plateau behavior is almost guaranteed by the 
functional form of the error function. However, one can pin down
the origin of plateau by looking closely at the late-time behavior
of the second term $\Tr(e^{\bt_1Q}\Pi e^{\bt_2Q}\Pi)$ in the connected correlator 
$Z_2(\bt_1,\bt_2)$ in \eqref{eq:corr-Pi}.
As we have seen in section \ref{sec:saddle}, this term can be evaluated by the saddle point approximation.
For $\bt_{1,2}=\bt\pm\ri t$, the saddle points are given by
\begin{equation}
\begin{aligned}
 \xi_{1,2}^*=\qu\arcsin(\la\pm\ri \tau)^2,
\end{aligned} 
\label{eq:saddle-SFF}
\end{equation}
and the saddle point value is given by
\begin{equation}
\begin{aligned}
 \Tr(e^{\bt_1Q}\Pi e^{\bt_2Q}\Pi)\sim \exp\left[\frac{\cF_0(\la+\ri\tau)+\cF_0(\la-\ri\tau)}{\hbar}\right].
\end{aligned} 
\end{equation}
This contribution decays exponentially at late-times and
the SFF approaches the plateau value given by the first term $\Tr(e^{(\bt_1+\bt_2)Q}\Pi)$
in \eqref{eq:corr-Pi}.
These saddle points \eqref{eq:saddle-SFF} can be thought of as the
eigenvalue instantons sitting at the complex conjugate pair of
points $E=-\xi^*_{1,2}$ and 
the transition from ramp to plateau is induced by the pair creation of
eigenvalue instantons as advocated in \cite{Okuyama:2018gfr}.

\begin{figure}[thb]
\centering
\includegraphics[width=8cm]{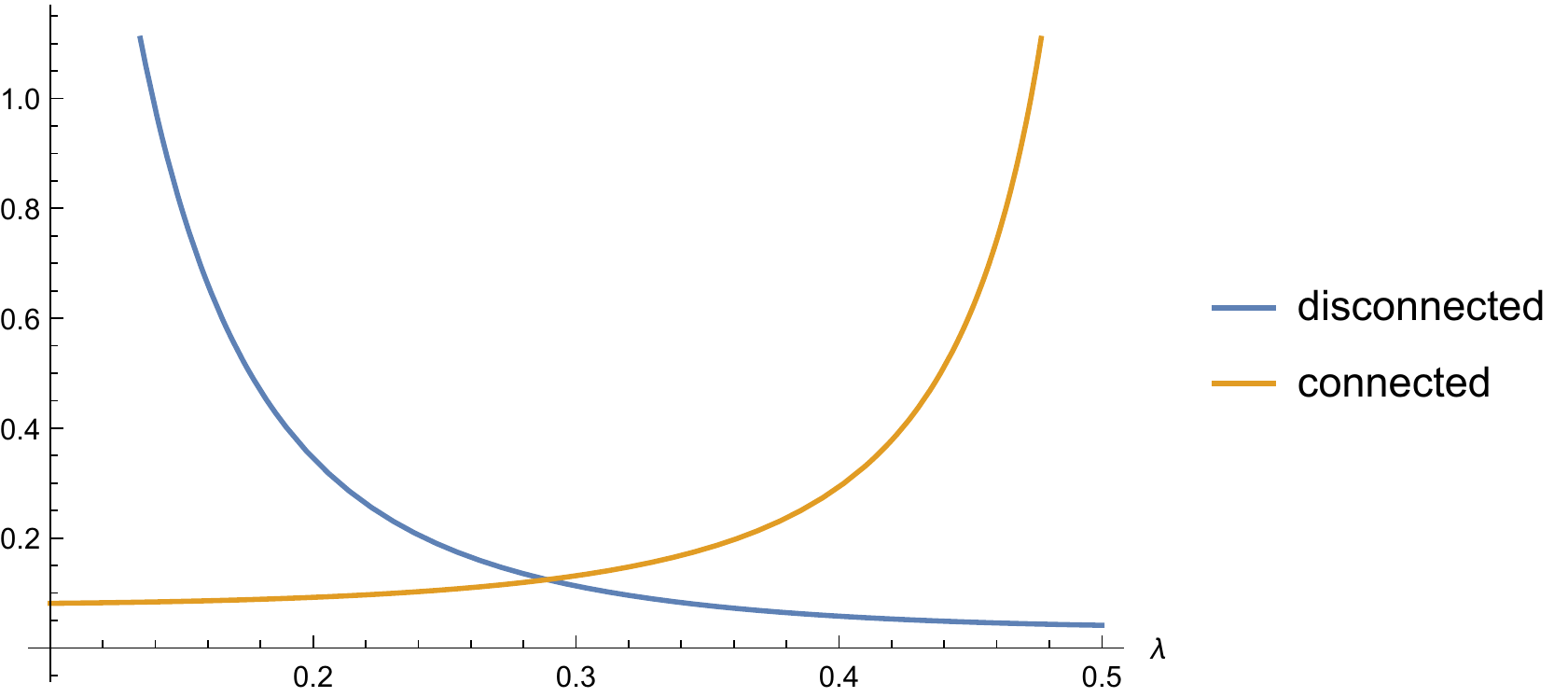}
  \caption{Plot of the disconnected part $\bra Z(\bt)\ket^2$ and
the connected part $\bra Z(\bt)^2\ket_{\conn}$ as a function of $\la=\hbar\bt$
at $\hbar=1/30$.} 
  \label{fig:exchange}
\end{figure}

Another interesting phenomenon is that the connected and the disconnected
contributions exchange dominance as we lower the temperature. This transition
is observed in a coupled SYK model \cite{Maldacena:2018lmt}
and it is expected to occur in JT gravity as well. To see this, 
let us compare the disconnected part $\bra Z(\bt)\ket^2$
and the connected part $\bra Z(\bt)^2\ket_{\conn}$ and
study their behavior as a function of $\bt$. Here
we set $\bt_1=\bt_2=\bt$ for simplicity. Since JT gravity
becomes a good approximation of the SYK model in the low energy limit, 
it is useful to study the behavior of two-boundary correlator 
in the 't Hooft limit.
At the leading order in the 't Hooft expansion we find
\begin{equation}
\begin{aligned}
\bra Z(\bt)\ket^2
&\approx \frac{\hbar}{4\pi (\arcsin\la)^3\sqrt{1-\la^2}}
e^{\frac{2\cF_0(\la)}{\hbar}},\\
\bra Z(\bt)^2\ket_{\conn}
&\approx \sqrt{\frac{\hbar}{4\pi(\arcsin 2\la)^3\sqrt{1-4\la^2}}}
e^{\frac{\cF_0(2\la)}{\hbar}}\text{Erf}
\Biggl(\rt{\frac{\cF_0(2\la)-2\cF_0(\la)}{\hbar}}\Biggr).
\end{aligned} 
\label{eq:con-dis}
\end{equation}
In Fig.~\ref{fig:exchange}, we show the plot of \eqref{eq:con-dis} for $\hbar=1/30$.
One can see that at high temperature
the disconnected part is dominant, but as we lower the temperature
the connected part becomes dominant below some critical temperature.
Thus we succeeded to reproduce the transition observed
in \cite{Maldacena:2018lmt} directly from the JT gravity computation.
In the bulk gravity picture, this is an analogue of the Hawking-Page transition
between two different topologies of spacetime.
At high temperature the two disconnected Euclidean black holes are dominant
while at low temperature the Euclidean wormhole connecting the two
boundaries becomes dominant. 

\section{Boundary creation operator and Hartle-Hawking state}\label{sec:boundary}

As we have seen in section \ref{sec:genus},
we can write the connected $n$-point amplitude as
\begin{equation}
\begin{aligned}
 \bra Z(\bt_1)\cdots Z(\bt_n)\ket_{\conn}\simeq B(\bt_1)\cdots B(\bt_n)F,
\end{aligned} 
\label{eq:B-corr}
\end{equation} 
where $F$ denotes the free energy \eqref{eq:genGF}
and the operator $B(\bt)$ is given by \eqref{eq:Bdef}.
$B(\bt)$ can be thought of as the ``boundary creation operator.''
The same operator has been considered in
the context of 2d gravity in \cite{Moore:1991ir}.
\eqref{eq:B-corr} should be understood as the equality up to the non-universal terms
at genus-zero for the one- and two-point functions,
which should be treated separately.

In a recent paper by Marolf and Maxfield
\cite{Marolf:2020xie},
the idea of boundary creation operators is also discussed. 
The important property of the boundary creation operators
is that they all commute and hence can be diagonalized simultaneously.
It is argued in \cite{Marolf:2020xie} that the simultaneous
eigenstate of the boundary creation operators, the so-called $\al$-state,
can be thought of as a member of an ensemble and the correlator
$\bra\prod_i Z(\bt_i)\ket$ is interpreted as the ensemble average.
Moreover, by reinterpreting
the earlier discussion of baby universes \cite{Coleman:1988cy,Giddings:1988cx,Giddings:1988wv}
from the viewpoint of AdS/CFT duality,
it is argued that
one can define the baby universe Hilbert space from the data of correlators
$\bra\prod_i Z(\bt_i)\ket$ and this Hilbert space 
includes many null states due to the 
bulk diffeomorphism invariance.\footnote{In a recent paper \cite{vafa},
it is argued that the baby universe Hilbert space must be one-dimensional
in a consistent quantum gravity on a spacetime with dimension $d>3$.}
To demonstrate these properties, a simple toy model is studied in \cite{Marolf:2020xie}
where the action $S$ of the model has only the topological term
given by the Euler characteristic $\chi$
of the 2d spacetime. 

In this section we will consider whether the proposal in \cite{Marolf:2020xie}
can be generalized to the JT gravity case.
Firstly, the boundary creation operator $B(\bt)$ defined in \eqref{eq:Bdef} clearly commutes
\begin{equation}
\begin{aligned}
 {[}B(\bt), B(\bt'){]}=0,
\end{aligned} 
\end{equation}
and hence one can try to diagonalize $B(\bt)$'s simultaneously. 
One immediate problem is that $B(\bt)$ in \eqref{eq:Bdef}
does not look like a hermitian operator, thus its eigenvalue
is not necessarily a real number.
According to the proposal in \cite{Marolf:2020xie},
this problem might be resolved on the physical Hilbert space, which is obtained 
by taking the quotient of the original Hilbert space
by the space of null states.
We do not have a clear understanding of how this happens.
In the rest of this section, we will examine
how the proposal of \cite{Marolf:2020xie} is generalized or modified 
in the case of JT gravity.

To study the proposal of \cite{Marolf:2020xie} in JT gravity,
it is convenient to use the free boson-fermion representation of 
the Witten-Kontsevich $\tau$-function 
(see e.g. \cite{BBT,Aganagic:2003qj,Kostov:2009nj,Kostov:2010nw} and references therein)
\begin{equation}
\begin{aligned}
\tau= e^F=\bra t|V\ket,
\end{aligned} 
\label{eq:tau-WK}
\end{equation}
where the state $\bra t|$ is given by the coherent state of free boson
\begin{equation}
\begin{aligned}
 \bra t|=\bra 0|\exp\left(\sum_{k=0}^\infty \frac{\til{t}_k\al_{2k+1}}{\gs(2k+1)!!}\right)
\end{aligned} 
\label{eq:bra-t}
\end{equation}
with $\al_n$ obeying the commutation relation $[\al_n,\al_m]=n\cob_{n+m,0}$.
$\til{t}_k$ in \eqref{eq:bra-t} is defined by
\begin{equation}
\begin{aligned}
 \til{t}_k=t_k-\cob_{k,1}.
\end{aligned} 
\end{equation}

To write down the state $|V\ket$ in \eqref{eq:tau-WK},
it is useful to introduce the free fermions $\psi_r,\psi_r^*~(r\in\mathbb{Z}+\hf)$
obeying the anti-commutation relation $\{\psi_r,\psi_s^*\}=\cob_{r+s,0}$.
Then $|V\ket$ is written as 
\begin{equation}
\begin{aligned}
 |V\ket=\exp\left(\sum_{m,n=0}^\infty A_{m,n}\psi_{-m-\hf}\psi^*_{-n-\hf}\right)|0\ket.
\end{aligned} 
\label{eq:V-ket}
\end{equation}
The generating function of $A_{m,n}$ for the Witten-Kontsevich $\tau$-function
is obtained in \cite{zhou2013explicit,zhou2015emergent,balogh2017geometric}.
The important property of the state $|V\ket$ is that
it satisfies the Virasoro constraint  \cite{Fukuma:1990jw,Dijkgraaf:1990rs}
\begin{equation}
\begin{aligned}
 \cL_n|V\ket=0\quad(n\geq-1),
\end{aligned} 
\label{eq:Vir-con}
\end{equation}
where the Virasoro generator $\cL_n$ is given by
\begin{equation}
\begin{aligned}
 \cL_n&=\qu\sum_{k\in\mathbb{Z}}:\al_{2k+1}\al_{2n-2k-1}:
-\frac{1}{2\gs}\al_{2n+3}+\frac{1}{16}\cob_{n,0}.
\end{aligned} 
\label{eq:Ln-def}
\end{equation}
%
In \cite{Sen:1990rz,Imbimbo:1990ua},
the Virasoro constraint of matrix model
is interpreted as the gauge symmetry of
closed string field theory in a minimal model background. 
This suggests that the Virasoro constraint 
is the analogue of the bulk diffeomorphism invariance discussed in \cite{Marolf:2020xie}.

Now let us consider the Hartle-Hawking state $|\text{HH}\ket$ \cite{Hartle:1983ai}. 
As discussed in \cite{Polchinski:1989fn},
it is natural to identify the Hartle-Hawking state $|\text{HH}\ket$
as ``the most symmetric state.'' In the present case,
$|V\ket$ is such a state since $|V\ket$ is invariant
under the Virasoro generators \eqref{eq:Ln-def}.
$|V\ket$ can be thought of as the $SL(2,\mathbb{C})$
invariant vacuum corresponding to the identity operator and it is a natural candidate
for the no-boundary state.
Thus we propose to identify the Hartle-Hawking state $|\text{HH}\ket$
with the state $|V\ket$ in \eqref{eq:V-ket}
\begin{equation}
\begin{aligned}
 |\text{HH}\ket=|V\ket.
\end{aligned} 
\label{eq:our-HH}
\end{equation}
In particular, this state satisfies the equation
$\cL_0|\text{HH}\ket=0$ which corresponds to 
the Wheeler-DeWitt equation.

Next we consider the interpretation of the correlator $\bra \prod_i Z(\bt_i)\ket$
in JT gravity. The correlator here refers to the full correlator
including both the connected and the disconnected parts.
One can generalize \eqref{eq:B-corr} to the full correlator
by acting $B(\bt)$'s on the $\tau$-function instead of the free energy
\begin{equation}
\begin{aligned}
 \bra Z(\bt_1)\cdots Z(\bt_n)\ket&\simeq\frac{B(\bt_1)\cdots B(\bt_n)\bra t|V\ket}
{\bra t|V\ket}\\
&=:
\frac{\bra t|\h{B}(\bt_1)\cdots \h{B}(\bt_n)|V\ket}{\bra t|V\ket},
\end{aligned} 
\end{equation}
where the operator $\h{B}(\bt)$ is written as
\begin{equation}
\begin{aligned}
 \h{B}(\bt)=\frac{1}{\rt{2\pi}}\sum_{n=0}^\infty \frac{\bt^{n+\hf}}{(2n+1)!!}\al_{2n+1}.
\end{aligned} 
\end{equation}
It turns out that the non-universal terms at genus-zero are correctly incorporated by
extending the summation to all $n\in\mathbb{Z}$. Namely we define
the operator $\h{Z}(\bt)$ by
\begin{equation}
\begin{aligned}
 \h{Z}(\bt)=\frac{1}{\rt{2\pi}}\sum_{n=-\infty}^\infty \frac{\bt^{n+\hf}}{(2n+1)!!}\al_{2n+1}.
\end{aligned} 
\label{eq:hat-Z}
\end{equation}
Then the full correlator is given by
\begin{equation}
\begin{aligned}
 \bra Z(\bt_1)\cdots Z(\bt_n)\ket=
\frac{\bra t|\h{Z}(\bt_1)\cdots \h{Z}(\bt_n)|\text{HH}\ket}{\bra t|\text{HH}\ket},
\end{aligned} 
\label{eq:HH-corr}
\end{equation}
where we used our identification $|\text{HH}\ket=|V\ket$.
To see that this is the correct prescription, 
let us consider the genus-zero part of
the one-point function
\begin{equation}
\begin{aligned}
\bra Z(\bt)\ket^{g=0} &=\frac{ \bra t| \h{Z}(\bt)|\text{HH}\ket}{\bra t|\text{HH}\ket} \Bigg|_{g=0}
=\bra 0|\exp\left(\sum_{k=1}^\infty \frac{\til{t}_k}{\gs (2k+1)!!}\al_{2k+1}\right)
\h{Z}(\bt)|0\ket\\
&=\frac{1}{\rt{2\pi}\gs}\sum_{k=1}^\infty \bt^{-k-\hf}\frac{\til{t}_k}{(2k-1)!!(-2k-1)!!}\\
&=\frac{e^{1/\bt}}{\rt{2\pi}\gs\bt^{3/2}}.
\end{aligned} 
\label{eq:g0-Z}
\end{equation} 
Here we have used $\til{t}_k=\frac{(-1)^k}{(k-1)!}~(k\geq1)$
and
\begin{equation}
\begin{aligned}
 (2k-1)!!(-2k-1)!!=(-1)^k.
\end{aligned} 
\label{eq:fac-rel}
\end{equation}
Similarly, the genus-zero part of the two-point function becomes
\begin{equation}
\begin{aligned}
 \bra Z(\bt_1)Z(\bt_2)\ket^{g=0}_{\conn}&=\frac{1}{2\pi}\Biggl\bra 0\Biggr|\sum_{k=0}^\infty
\frac{\bt_1^{k+\hf}}{(2k+1)!!}\al_{2k+1}
\sum_{n=0}^\infty
\frac{\bt_2^{-n-\hf}}{(-2n-1)!!}\al_{-2n-1}\Biggl|0\Biggr\ket\\
&=\frac{1}{2\pi}\sum_{n=0}^\infty \frac{\bt_1^{n+\hf}\bt_2^{-n-\hf}}{(2n-1)!!(-2n-1)!!}
\\
&=\frac{\rt{\bt_1\bt_2}}{2\pi(\bt_1+\bt_2)}.
\end{aligned} 
\label{eq:g0-ZZ}
\end{equation}
\eqref{eq:g0-Z} and \eqref{eq:g0-ZZ} agree with the known result of the genus-zero part
in JT gravity.
Using the relation \eqref{eq:fac-rel} one can show that $\h{Z}(\bt)$'s commute at least formally
\begin{equation}
\begin{aligned}
 {[}\h{Z}(\bt_1),\h{Z}(\bt_2){]}
&=\frac{1}{2\pi}\sum_{n,k\geq0}\left[\frac{\bt_1^{k+\hf}}{(2k+1)!!}\al_{2k+1},
\frac{\bt_2^{-n-\hf}}{(-2n-1)!!}\al_{-2n-1}\right]\\
&+\frac{1}{2\pi}\sum_{n,k\geq0}
\left[\frac{\bt_1^{-k-\hf}}{(-2k-1)!!}\al_{-2k-1},
\frac{\bt_2^{n+\hf}}{(2n+1)!!}\al_{2n+1}\right]\\
&=\frac{1}{2\pi}\sum_{n\geq0}(-1)^n\Bigl(\bt_1^{n+\hf}\bt_2^{-n-\hf}-\bt_1^{-n-\hf}\bt_2^{n+\hf}\Bigr)\\
&=\frac{\rt{\bt_1\bt_2}}{2\pi}\left(\frac{1}{\bt_1+\bt_2}-\frac{1}{\bt_1+\bt_2}\right)\\
&=0.
\end{aligned} 
\end{equation}

Our proposal \eqref{eq:HH-corr} is 
consistent with the identification
of the one-point function $\bra Z(\bt)\ket$
as the wavefunction of the Hartle-Hawking state, which is usually 
assumed in 2d gravity literature
(see e.g.~\cite{Ginsparg:1993is} and references therein)
\begin{equation}
\begin{aligned}
 \bra Z(\bt)\ket=\Psi_{\text{HH}}(\bt)=\bra \bt|\text{HH}\ket,
\end{aligned} 
\end{equation}
where $\bra \bt|$ is given by
\begin{equation}
\begin{aligned}
 \bra \bt|=\frac{\bra t|\h{Z}(\bt)}{\bra t|\text{HH}\ket}.
\end{aligned} 
\end{equation}
More generally, the multi-point correlator is written as
\begin{equation}
\begin{aligned}
\bra Z(\bt_1)\cdots Z(\bt_n)\ket&=\bra \bt_1,\cdots,\bt_n|\text{HH}\ket,\\
\bra \bt_1,\cdots,\bt_n|&=\frac{\bra t|\h{Z}(\bt_1)\cdots \h{Z}(\bt_n)}{\bra t|\text{HH}\ket}.
\end{aligned} 
\end{equation}

Our expression \eqref{eq:HH-corr} is different from the proposal in \cite{Marolf:2020xie}
\begin{equation}
\begin{aligned}
 \bra Z(\bt_1)\cdots Z(\bt_n)\ket=
\frac{\bra \text{HH}|\h{Z}(\bt_1)\cdots \h{Z}(\bt_n)|\text{HH}\ket}{\bra\text{HH}|\text{HH}\ket}.
\end{aligned}
\label{eq:MM-prop} 
\end{equation}
This difference comes from the fact that
the bra and the ket are treated asymmetrically in the free boson/fermion 
representation of the $\tau$-function \eqref{eq:tau-WK}.
In other words, our expression \eqref{eq:HH-corr} corresponds to a special
(Euclidean) time-slicing of the spacetime where the initial state has no boundary 
and all the boundaries are on the final state. 
At present, it is not clear to us
how to reconcile our \eqref{eq:HH-corr} and the proposal \eqref{eq:MM-prop} 
in \cite{Marolf:2020xie}.

\section{Conclusions and outlook}\label{sec:conclusion}

We have studied the multi-boundary correlators in JT gravity
using the KdV constraints obeyed by these correlators.
Along the way, we have defined the off-shell generalization of the effective potential and
have studied the WKB expansion of the Baker-Akhiezer functions as well.
In particular, we have computed the genus expansion of the connected
two-boundary correlator $\bra Z(\bt_1)Z(\bt_2)\ket_\conn$
as well as its low temperature expansion. We have found
that the two-point function is written in terms of the error function and
the ramp and plateau behavior of the SFF in JT gravity is 
explained by the functional form of this error function.
We have also confirmed the picture put forward in \cite{Okuyama:2018gfr}
that the transition from ramp to plateau is induced by the pair creation of eigenvalue
instantons. 

There are many interesting open questions.
In section \ref{sec:boundary} we briefly discussed a possible connection to the recent work by
Marolf and Maxfield \cite{Marolf:2020xie} which clearly deserves further
investigation. It would be interesting to
construct the $\al$-state which simultaneously diagonalizes
the operator $\h{Z}(\bt)$ in \eqref{eq:hat-Z} 
and see how the argument in \cite{Marolf:2020xie}
is generalized to the JT gravity case.
In particular, it is interesting to see what
the non-factorized contribution $\bra Z(\bt_1)Z(\bt_2)\ket_\conn$ coming from
the Euclidean wormhole \cite{Maldacena:2004rf,ArkaniHamed:2007js}
looks like in the $\al$-state.
The pure topological gravity would be a good starting point to study
this problem since the explicit form of the $n$-point correlator is known 
in the literature \cite{okounkov2002generating,buryak,Alexandrov:2019eah}.

It is emphasized in \cite{Marolf:2020xie} that non-perturbative effects
are important to realize
the massive truncation of the Hilbert space 
by the diffeomorphism invariance.
The free fermion representation of the state $|V\ket$ in \eqref{eq:V-ket} is defined by
the asymptotic expansion in $\gs$ and hence
it only makes sense as a perturbative expansion.
However, it is possible to include the
effect of D-instanton corrections systematically within this 
framework \cite{Fukuma:1996hj,Fukuma:1996bq,Fukuma:1999tj}.
It would be interesting to study such D-instanton effects in JT gravity
and see how they affect the argument of diffeomorphism invariance
in JT gravity.

In \cite{Penington:2019kki,Almheiri:2019qdq} it is
argued that the Page curve
for the black hole evaporation is correctly reproduced if we include the
contribution of replica wormholes in the computation of entropy
of Hawking radiation
using the replica method in the gravity path integral.
One can immediately apply our formalism to compute the contribution of
the replica wormholes in pure JT gravity sector.
To model the black hole microstates one can add the end of the world (EOW) branes
to JT gravity \cite{Penington:2019kki,Marolf:2020xie}.
It would be interesting to construct a generalization of the JT gravity matrix model
which incorporates the degrees of freedom of the EOW branes.

As discussed in \cite{Maldacena:2019cbz,Cotler:2019nbi}, the matrix model description of JT gravity
can be generalized to the 2d de Sitter space
by analytically continuing the boundary length $\bt$ to imaginary value $\bt\to\pm\ri\ell$.
In \cite{Cotler:2019dcj} the boundary creation/annihilation operators are
considered in this de Sitter setting. It would be interesting to see 
how they are related to our discussion in section \ref{sec:boundary}.

Finally, it would be interesting to generalize
our computation in this paper to JT supergravity \cite{Stanford:2019vob}.
In particular, the genus expansion of
JT supergravity on orientable surfaces without time-reversal symmetry
can be computed from the Brezin-Gross-Witten $\tau$-function \cite{norbury}.
We will report on the computation of JT supergravity case elsewhere.

\acknowledgments
This work was supported in part by JSPS KAKENHI Grant
Nos.~19K03845 and 19K03856,
and JSPS Japan-Russia Research Cooperative Program.

\appendix

\section{Wavefunction of microscopic loop operators}\label{sec:micro}

In this appendix we will consider the correlator of microscopic loop operators in
the presence of one macroscopic loop operator. It is easily
obtained 
by differentiating $\bra Z(\bt)\ket$
\begin{equation}
\begin{aligned}
 \del_{n_1}\del_{n_2}\cdots \bra Z(\bt)\ket=\bra \tau_{n_1}\tau_{n_2}\cdots Z(\bt)\ket.
\end{aligned} 
\end{equation}
It is convenient to define the normalized correlator
\begin{equation}
\begin{aligned}
 \big\bra\!\big\bra\prod_i \tau_{n_i}\big\ket\!\big\ket:
=\frac{\bra \prod_i \tau_{n_i} Z(\bt)\ket}{\bra Z(\bt)\ket},
\end{aligned} 
\end{equation}
which can 
be thought of as the wavefunction of microscopic loop operators \cite{Moore:1991ir,Ginsparg:1993is}.

For instance, the one-point function $\big\bra\!\big\bra\tau_{n}\big\ket\!\big\ket$
at the leading order is given by
\begin{equation}
\begin{aligned}
 \big\bra\!\big\bra\tau_{n}\big\ket\!\big\ket=\del_n\log\bra Z(\bt)\ket\approx 
\frac{1}{\hbar}\del_n \cF_0(\la).
\end{aligned} 
\label{eq:tau-dF}
\end{equation}
The derivative of $\cF_0(\la)$ with respect to the coupling $t_n$ can be computed by using
the fact
that $V_{\text{eff}}(\xi)$ and $\cF_0(\la)$ are related by the Legendre
transformation.
Thus we find
\begin{equation}
\begin{aligned}
 \frac{\del\cF_0(\la)}{\del t_n}\Bigg|_{\la~\text{fixed}}&=\frac{\del}{\del t_n}\Bigg|_{\la~\text{fixed}}
\Bigl(\la\xi_*-V_{\text{eff}}(\xi_*)\Bigr)\\
&=\la\frac{\del\xi_*}{\del t_n}
-V_{\text{eff}}'(\xi_*)\frac{\del\xi_*}{\del t_n}-\frac{\del V_{\text{eff}}(\xi_*)}{\del t_n}\Bigg|_{\xi_*~\text{fixed}}\\
&=-\frac{\del V_{\text{eff}}(\xi_*)}{\del t_n}\Bigg|_{\xi_*~\text{fixed}}.
\end{aligned} 
\label{eq:FV-Legendre}
\end{equation}
In the last step we have used the saddle point equation $\la=V_{\text{eff}}'(\xi_*)$. 
From the explicit form of the off-shell effective potential in \eqref{eq:off-Veff},
one can easily compute the derivative $-\del_n V_{\text{eff}}(\xi_*)$.
From \eqref{eq:tau-dF} and \eqref{eq:FV-Legendre}, 
for the on-shell JT gravity case $t_n=\ga_n$ we find
the wavefunction of the microscopic loop operator $\tau_n$ at the leading order
in the 't Hooft expansion \eqref{eq:tHooft}
\begin{equation}
\begin{aligned}
 \big\bra\!\big\bra\tau_{n}\big\ket\!\big\ket=
\frac{\arcsin(\la)^{2n+1}}{ 2^{n}(2n+1)!!\hbar}+\cO(\hbar^0).
\end{aligned} 
\end{equation} 
It turns out that the wavefunction is factorized at the leading order 
in the 't Hooft expansion
\begin{equation}
\begin{aligned}
 \big\bra\!\big\bra\prod_i \tau_{n_i}\big\ket\!\big\ket&\approx e^{-\frac{\cF_0(\la)}{\hbar}}\prod_i\del_{n_i}e^{\frac{\cF_0(\la)}{\hbar}}\\
&\approx\prod_i \frac{\del_{n_i}\cF_0(\la)}{\hbar}\\
&\approx \prod_i \big\bra\!\big\bra\tau_{n_i}\big\ket\!\big\ket.
\end{aligned} 
\end{equation}

One can go beyond the leading order and compute the
higher order correction to the wavefunction of microscopic loop operators
by using the off-shell generalization of the free energy $\cF$ in \eqref{eq:offFnexp}.
After some algebra, 
we find the first order correction to the 't Hooft expansion
\begin{equation}
\begin{aligned}
\big\bra\!\big\bra\tau_0^k\prod_{i=1}^m \tau_{n_i}\big\ket\!\big\ket
=&\frac{\arcsin(\la)^k}{\hbar^{m+k}}
\prod_{i=1}^m\frac{\arcsin(\la)^{2n_i+1}}{ 2^{n_i}(2n_i+1)!!}\\
\times &\Biggl[1+\hbar\biggl(2\sum_{i=1}^mn_i+m+k\biggr)\left(
\frac{2\sum_{i=1}^mn_i+m+k-5}{\rt{1-\la^2}\arcsin(\la)^3}+\frac{\la}{(1-\la^2)\arcsin(\la)^2}\right)\\
&\qquad+\hbar\frac{5k-k^2}{\arcsin(\la)^3}+\cO(\hbar^2)\Biggr],
\end{aligned} 
\label{eq:macro-micro}
\end{equation}
where $n_i>0~(i=1,\cdots,m)$.


\bibliography{paper}
\bibliographystyle{utphys}
\end{document}